\documentclass[preprint,notoc,float]{JHEP3}

\usepackage{epsfig}
\def\eslt{\not\!\!{E_T}}

\def\to{\rightarrow}

\def\bi{\begin{itemize}}
\def\ei{\end{itemize}}
\def\be{\begin{equation}}
\def\ee{\end{equation}}
\def\bea{\begin{eqnarray}}
\def\eea{\end{eqnarray}}
\def\te{\tilde e}

\def\tu{\tilde u}

\def\tb{\tilde b}

\def\td{\tilde d}

\def\tst{\tilde t}
\def\ttau{\tilde \tau}
\def\tmu{\tilde \mu}
\def\tg{\tilde g}
\def\tnu{\tilde\nu}
\def\tell{\tilde\ell}
\def\tq{\tilde q}
\def\tw{\widetilde W}
\def\tz{\widetilde Z}
\def\alt{\stackrel{<}{\sim}}

\title{
SUSY Normal Scalar Mass Hierarchy Reconciles $(g-2)_\mu$, 
$b\to s\gamma$ and Relic Density
}

\author{Howard Baer, Alexander Belyaev, Tadas Krupovnickas
and Azar Mustafayev
\\ Department of Physics, Florida State University\\ 
Tallahassee, FL 32306, USA\\
E-mail: \email{baer@hep.fsu.edu}, \email{belyaev@hep.fsu.edu}, 
\email{tadas@hep.fsu.edu}, \email{mazar@hep.fsu.edu}}

\preprint{\vbox{\hbox{FSU-HEP-031104}}} 

\abstract{
Recent experimental and theoretical determinations of $(g-2)_\mu$,
$\Omega_{CDM}h^2$ and $BF(b\to s\gamma )$ place exceedingly tight 
constraints on the minimal supergravity (mSUGRA) model. We 
advocate relaxing the generational universality of mSUGRA, so that
GUT scale third generation scalar masses are greater than the 
(degenerate) first and second generation scalar masses 
(a normal scalar mass hierarchy (NMH)). The non-degeneracy
allows for a reconciliation of all the above constraints, and
also respects FCNC limits from $B_d-\overline{B}_d$ mixing
and $b\to s\gamma$.
The NMH SUGRA model leads to the prediction of relatively light
first and second generation sleptons. This yields 
large rates for multilepton 
collider signatures at the CERN LHC and also possibly at the
Fermilab Tevatron. The spectrum of light sleptons should be accessible
to a $\sqrt{s}=0.5-1$ TeV linear $e^+e^-$ collider.
}


\begin{document}

\section{Introduction}
\label{sec:intro}

The minimal supergravity model (mSUGRA) has long served as the paradigm
supersymmetric model for testing the phenomenological consequences of
weak scale supersymmetry\cite{msugra}. 
The model is constructed by starting with the
general supergravity Lagrangian of $N=1$ supersymmetric 
gauge theories\cite{cremmer}. The gauge symmetry and matter content
of the Standard Model are assumed, where now matter fermions 
comprise just one component of left chiral superfields, and the Higgs sector
must be expanded by an additional doublet to cancel triangle anomalies.
Supersymmetry breaking occurs within a postulated hidden sector,
and can lead to weak scale soft SUSY breaking terms of mass
$\tilde m =\Lambda^2/M_{Pl}$ if the SUSY breaking parameter 
$\Lambda\sim 10^{11}$ GeV. Motivated by the degeneracy solution to the SUSY
flavor and CP problems\cite{dg}, the simple choice of a flat K\"ahler metric
can be made. This leads at tree level to degenerate scalar masses at 
some high mass scale, 
usually taken to be $M_{GUT}\sim 2\times 10^{16}$ GeV. 
A simple choice of the gauge kinetic function leads also to gaugino 
mass unification. Taking the limit as $M_{Pl}\to\infty$ while
keeping $m_{3/2}$ fixed leads to a globally supersymmetric 
renormalizable gauge theory where supersymmetry
is broken by the presence of soft SUSY 
breaking terms (the MSSM). The weak scale parameters are related to
those of the GUT scale by RG evolution, and electroweak symmetry 
is broken radiatively.
The weak scale sparticle masses and mixings are determined by the
parameters 
\be
m_0,\ m_{1/2},\ A_0,\ \tan\beta,\ sign (\mu ) ,
\ee
where $m_0$ is the common scalar mass, $m_{1/2}$ is the common gaugino mass,
and $A_0$ is the common trilinear soft SUSY breaking term all
valid at $Q=M_{GUT}$. Throughout this work, we will adopt the value
$m_t=175$ GeV.

The mSUGRA model as defined above determines the large number of MSSM
parameters in terms of just a few parameters stipulated at $Q=M_{GUT}$, 
and therefore provides a very rich phenomenology in terms of just a few
free parameters. However, the model has long been susceptible to 
criticism owing primarily to the assumption of universality, especially 
amongst scalar masses\cite{sug_crit}. 
First, there is no theoretical reason for choosing a flat
K\"ahler metric, and in general the scalar masses can be arbitrary.
The arbitrary soft breaking masses in general lead to violation of 
constraints from FCNC processes\cite{fcnc}, 
and complex soft breaking parameters
can lead to CP violating phenomena at large rates\cite{masiero}. 
It has been shown that even if
universality is imposed at tree level, the universality will be broken by
loop corrections\cite{sugra_loop}. 
We note here that there do exist symmetries which
guarantee universality of soft breaking terms {\it within} a 
generation, for instance, $SO(10)$ gauge symmetry. But then the 
non-universality would still be manifest as a splitting amongst the 
generations, and with the Higgs boson soft masses.

In recent years, supersymmetric models have become increasingly constrained
by a variety of measurements~\cite{constraints}. 
These include determination of the
branching fraction 
$BF(b\to s\gamma )$~\cite{belle,cleo,aleph,gambino,bsg,bsg_bb}, 
the muon anomalous
magnetic moment $a_\mu =(g-2)_\mu/2$\cite{e821} and the relic density of
cold dark matter in the universe. The most recent development comes from the
Muon $g-2$ Collaboration, which published results on $(g-2)_\mu$ for
the negative muon along with earlier results on the positive muon.
In addition, theoretical determinations of $(g-2)_\mu$ have been
presented by Davier {\it et al.}\cite{davier} and 
Hagiwara {\it et al.}\cite{kaoru} which use recent data on 
$e^+e^-\to hadrons$ at low energy
to determine the hadronic vacuum polarization contribution to the
muon magnetic moment. Combining the latest experiment and theory
numbers, we find the deviation of $a_\mu$ to be:
\bea
\Delta a_\mu &=&(27.1\pm 9.4)\times 10^{-10}\ \ \ ({\rm Davier}\ et\ al.)\\
\Delta a_\mu &=&(31.7\pm 9.5)\times 10^{-10}\ \ \ ({\rm Hagiwara}\ et\ al.) .
\label{eq:gm2_h}
\eea
The Davier {\it et al.} group also presents a number using $\tau$ decay data to
determine the hadronic vacuum polarization, which gives
$\Delta a_\mu =(12.4\pm 8.3)\times 10^{-10}$, {\it i.e.} nearly consistent
with the SM prediction. However, there seems to be growing consensus
that the numbers using the $e^+e^-$ data are more to be trusted, since 
they offer a direct determination of the hadronic vacuum polarization.
The $\sim 3\sigma$ deviation in $a_\mu$ using the $e^+e^-$ data 
can be explained in a supersymmetric
context if second generation sleptons (smuons and mu sneutrinos)
and charginos and neutralinos are relatively light\cite{gm2_susy}.

In SUSY models with minimal flavor violation, the flavor changing decay
$b\to s\gamma$ can still occur. The decay proceeds via Feynman graphs 
including $\tst_{1,2}\tw_{1,2}$ and $tH^+$ loops, in addition to the
SM contribution from a $tW$ loop. 
The branching fraction $BF(b\to s\gamma )$ has recently been measured by
the BELLE\cite{belle}, CLEO\cite{cleo} and ALEPH\cite{aleph}
collaborations.  Combining statistical and systematic errors in
quadrature, these measurements give $(3.36\pm 0.67)\times 10^{-4}$
(BELLE), $(3.21\pm 0.51)\times 10^{-4}$ (CLEO) and $(3.11\pm 1.07)\times
10^{-4}$ (ALEPH). A weighted averaging of these results yields 
$BF(b\to s\gamma )=(3.25\pm 0.37) \times 10^{-4}$.
To this we should add uncertainty
in the theoretical evaluation, which within the SM dominantly comes from
the scale uncertainty, and is about 12\%.\footnote{We caution the reader
that the SUSY contribution may have a larger theoretical uncertainty,
particularly if $\tan\beta$ is large. An additional theoretical
uncertainty that may increase the branching ratio in the SM is pointed
out in Ref. \cite{gambino}.} Experiment and theory 
together imply the bound,
\be
BF(b\to s\gamma )= (3.25\pm 0.54) \times 10^{-4} .
\ee
The calculation of SM and supersymmetric contributions to 
$BF(b\to s\gamma )$ used here is based upon the
program of Ref. \cite{bsg_bb}. Since the SM value of $BF(b\to s\gamma )$
is relatively close to the central measured value, rather large
values of $m_{\tst_{1,2}}$, $m_{\tw_{1,2}}$ and $m_{H^+}$ are expected,
which help to suppress the SUSY loop contributions to the decay amplitude.

Finally, the recent precision mapping of anisotropies in the 
cosmic microwave background radiation by the WMAP collaboration
have fitted the relic density of cold dark matter (CDM) in a $\Lambda CDM$ 
universe\cite{wmap,wmap_rev}. 
The constraint on the relic density of neutralinos
produced in the early universe is then
\be
\Omega_{\tz_1}h^2 =0.1126\pm 0.0081 .
\label{eq:oh2}
\ee
The new relic density constraint is highly restrictive. In the mSUGRA 
model, only a few regions give rise to this narrow band of $\Omega_{\tz_1}h^2$:
\begin{itemize}
\item The bulk region at low $m_0$ and low $m_{1/2}$, where neutralino
annihilation in the early universe proceeds dominantly via $t$-channel
slepton exchange. This region is now largely excluded 
(save where it overlaps with the stau co-annihilation region) because the WMAP
allowed region gives rise to low values of $m_h$ in conflict with
LEP2 bounds, and also large deviations in the $BF(b\to s\gamma )$.
\item The stau co-annihilation region at low $m_0$ where
$m_{\ttau_1}\simeq m_{\tz_1}$, so that $\tz_1 -\ttau_1$ and
$\ttau_1^+ -\ttau_1^-$ co-annihilation provides a ready sink for
neutralinos in the early universe\cite{stau}.
\item The hyperbolic branch/focus point region (HB/FP) at large
$m_0$ near the boundary of parameter space where $\mu^2$ 
becomes small\cite{hb_fp}.
The $\tz_1$ becomes increasingly higgsino-like in this region, 
facilitating its annihilation rate into $WW$, $ZZ$ and $Zh$
in the early universe.
\item The $A$-resonance annihilation region at large $\tan\beta$,
where $2m_{\tz_1}\sim m_A,\ m_H$. In this region, the $A$ and $H$
widths can be quite broad ($\sim 10-40$ GeV), giving rise to a relatively broad
$A$- annihilation ``funnel''\cite{Afunnel}.
\end{itemize}
In addition, other regions can occur such as the $h$ resonance annihilation 
region (a very narrow strip at low  $m_{1/2}$ where $2m_{\tz_1}\simeq m_h$)
and stop co-annihilation region (at the edge of parameter space for very 
particular values of $A_0$ parameter).

A recent $\chi^2$ analysis of $\Omega_{\tz_1}h^2$, $\Delta a_\mu$
and $BF(b\to s\gamma )$ has been made\cite{sug_chi2}, and has found
the HB/FP, $A$-funnel and stau regions to coincide with all 
experimental constraints. The analysis of Ref. \cite{sug_chi2}
pre-dated the new experimental and theoretical determinations of
$\Delta a_\mu$\footnote{Explictly, the theory calculation of 
Narison\cite{narison} using $e^+e^-\to hadrons$ data was used,
which gave $\Delta a_\mu =(24.1\pm 14)\times 10^{-10}$, {\it i.e} just a 
$1.7\sigma$ deviation from the SM prediction.}. The improved
$\Delta a_\mu$ numbers, as determined experimentally by E821
and theoretically by Davier {\it et al.} and by Hagiwara {\it et al.}, 
point to a larger
deviation from the SM, and in particular, to relatively light second generation
slepton masses, which do not occur in the HB/FP region, or in the
large $m_{1/2}$ part of the $A$-funnel or stau co-annihilation
region of the mSUGRA model.

Our goal in this paper is first to present updated $\chi^2$ fits to
the mSUGRA model, including the new $\Delta a_\mu$ determination.
While the new $\Delta a_\mu$ numbers prefer lighter sleptons, the
$BF(b\to s\gamma )$ numbers prefer heavy squarks,
typically in the several TeV range. The tension between these two 
constraints, coupled with the tight $\Omega_{\tz_1}h^2$ limits from WMAP,
leave only small viable portions of mSUGRA parameter space. 
We present these
results in Sec. \ref{sec:msugra}.

In Sec. \ref{sec:nmh}, we advocate relaxing the universality assumption
of the mSUGRA model.
In this case, constraints from FCNC processes must be addressed.
The FCNC constraints apply most strongly to first and second generation
scalar masses. 
We explore a scenario in  which first and
second generation scalars remain degenerate, while
allowing for a significant splitting with third generation
scalars. In this case, heavy (multi-TeV) third generation
scalars are preferred by $BF(b\to s\gamma )$ constraints,
while rather light first and second  generation scalars are
preferred by $\Delta a_\mu$.
The scenario is called the
normal scalar mass hierarchy (NMH), to distinguish it from
earlier studies which advocated an inverted scalar mass 
hierarchy (IMH)\cite{imh}.
The scenario respects constraints from $B_d-\overline{B}_d$ mixing.
The resulting model parameter space only allows low values of $m_{1/2}$, 
save for a small portion of the HB/FP region. Significant regions of 
parameter space survive all three constraints as given above. 
Light first and second generation sleptons of mass just a few
hundred GeV are characteristic to this scenario.

In Sec. \ref{sec:collider}, we examine consequences of the NMH scenario for
collider experiments.
When first and second generation sleptons become light, then
chargino and neutralino decays to electrons and muons  are
enhanced, leading  to large rates for distinctive multilepton
plus jets plus $\eslt$ events at the CERN LHC, and possibly
observable trilepton rates at the  Fermilab Tevatron. At a
linear $e^+e^-$ collider operating at $\sqrt{s}=0.5-1$ TeV,
first and second generation sleptons should be  accessible to
discovery, while tau sleptons would likely be beyond reach.

We summarize our results in Sec. \ref{sec:conclusions}.

\section{$\chi^2$ fit of mSUGRA model}
\label{sec:msugra}

Our first objective is to calculate the $\chi^2$ quantity formed from
$\Delta a_\mu$, $BF(b\to s\gamma )$ and $\Omega_{\tz_1}h^2$, as given in 
Eq'ns (\ref{eq:gm2_h}-\ref{eq:oh2}).
For $\Omega_{\tz_1}h^2$, 
we only use the upper bound contribution to the $\chi^2$, since
only the upper bound from WMAP is strictly applicable, {\it i.e.} there
may be other forms of dark matter in the universe. In addition, we
use the SM $a_\mu$ determination as given by Hagiwara {\it et al.}, 
which includes the latest fit to the hadronic vacuum polarization
from $e^+e^-\to hadrons$ at low energy. We generate the mSUGRA model
particle spectrum using Isajet 7.69\cite{isajet}, which includes two loop RGE
running of all couplings and soft SUSY breaking terms, 
minimizes the one loop effective potential at an optimized scale 
choice (which accounts for leading two loop terms), 
and which implements the complete set of log and finite 
corrections to all sparticle masses\footnote{Good agreement is found between 
Isajet and programs such as Suspect, SoftSUSY and Spheno in most of 
the mSUGRA model parameter space\cite{kraml}}.

Once the superparticle mass spectrum and mixings are generated, we calculate
the quantity $\Omega_{\tz_1}h^2$ using the Isared program, which includes
relativistic thermal averaging\cite{gondolo} 
of all relevant neutralino annihilation
and co-annihilation processes as calculated exactly at tree level using 
the CompHEP program\cite{relic}. We evaluate $\Delta a_\mu$ as in 
Ref.~\cite{gm2_susy}, 
and the $BF(b\to s\gamma )$ as outlined in Ref. \cite{bsg_bb},
which uses a tower of effective theories approach\cite{anlauf} 
and includes NLO corrections
to the running of Wilson coefficients between $Q=M_W$ and 
$Q=m_b$\cite{misiak}, 
and also includes the full set of NLO QCD corrections to the 
$b\to s\gamma$ process at the scale $Q=m_b$\cite{greub}. In addition, in our 
calculations, we implement the running value of the $b$ Yukawa coupling 
including MSSM threshold corrections for calculations 
above the scale $Q=M_W$\cite{carena}.
Our $BF(b\to s\gamma )$ results then agree well 
with those of Ref. \cite{giudice} at 
both small and large $\tan\beta$.

For every point in mSUGRA model parameter space, we calculate the $\chi^2$
quantity formed from $\Delta a_\mu$, $BF(b\to s\gamma )$ and 
$\Omega_{\tz_1}h^2$. For convenience, we plot in the $m_0\ vs.\ m_{1/2}$
plane the quantity $\sqrt{\chi^2}$, which is color coded according to 
the legend shown in Fig. \ref{fig:msugra}: {\it i.e.} green regions have
relatively low $\chi^2$, 
while red has a high $\chi^2$ and yellow is intermediate.
For variations in the two degrees of freedom ($m_0$ and $m_{1/2}$),
$\chi^2=2.3$ ($\sqrt{\chi^2}=1.52$) corresponds to a 
$1\sigma$ deviation from the 
combined measurements, $\chi^2=6.18$ ($\sqrt{\chi^2}=2.49$) corresponds to a 
$2\sigma$ deviation and  $\chi^2=11.83$ ($\sqrt{\chi^2}=3.44$) corresponds 
to a $3\sigma$ deviation.

Our first results are shown in Fig. \ref{fig:msugra}, where we plot for
$A_0=0$, $\mu >0$ and $\tan\beta$ values of 10, 30, 55 and 58.
For $\mu <0$, $\chi^2$ fits are almost always worse, since $\Delta a_\mu$
favors $\mu >0$. Variations in the $A_0$ parameter typically lead to 
no qualititative changes, save for those special $A_0$ values 
which drive the top squark mass to very low values.
The gray regions are excluded theoretically, while the blue regions 
are excluded by LEP2 searches for chargino pair production.
In Fig. \ref{fig:msugrac}, we
present contour plots for $BF(b\to s\gamma)$, $a_\mu$ and $\Omega h^2$
for the same parameters as Fig.~\ref{fig:msugra}; these figures
complement those of Fig. \ref{fig:msugra} and explain the 
details of the $\chi^2$ behavior.

\FIGURE{%
\hspace*{-0.5cm}
\epsfig{file=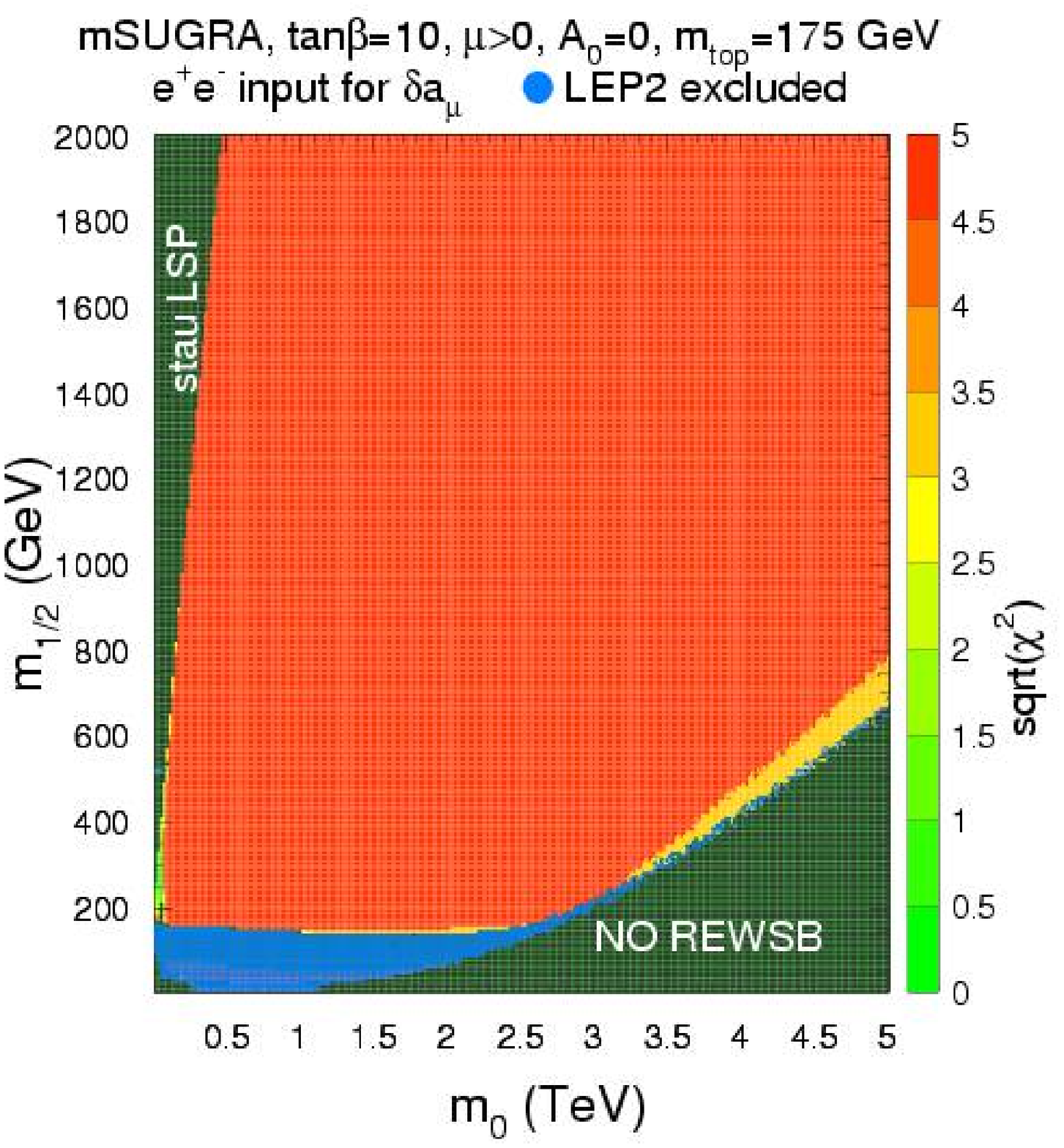,width=7.5cm}
\hspace*{-0.5cm}
\epsfig{file=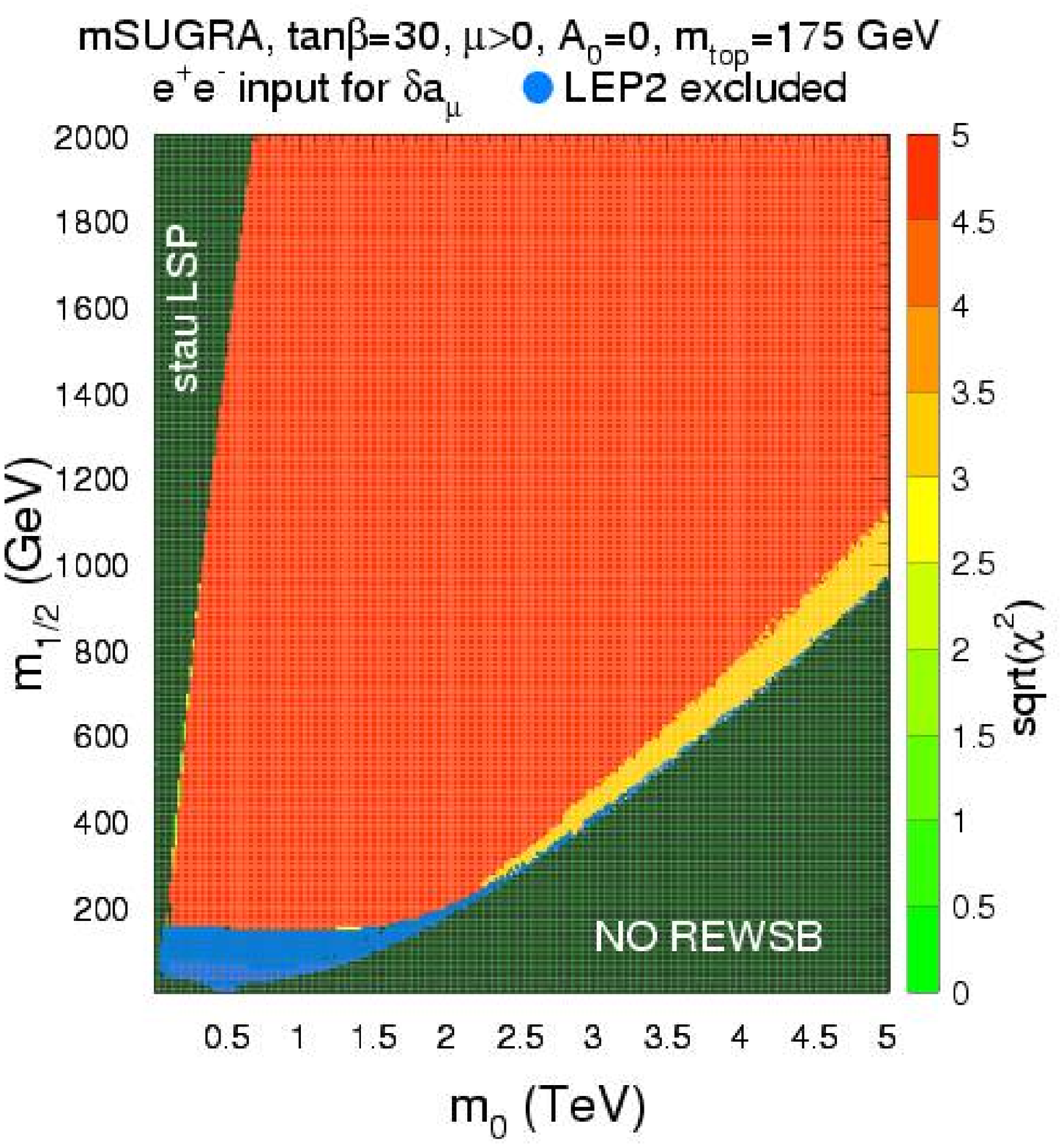,width=7.5cm}
\hspace*{-0.5cm}
\\
\hspace*{-0.5cm}
\epsfig{file=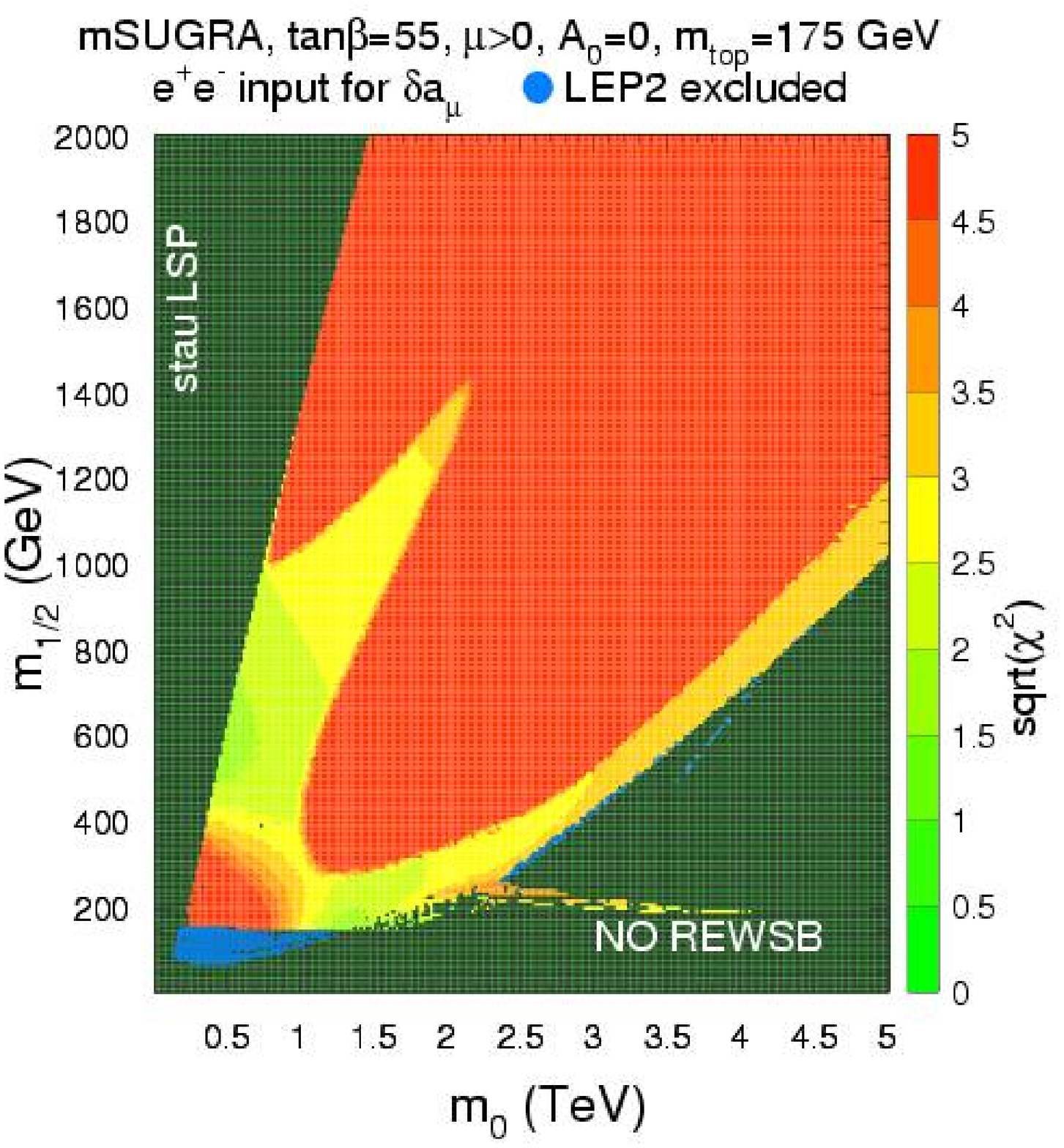,width=7.5cm}
\hspace*{-0.5cm}
\epsfig{file=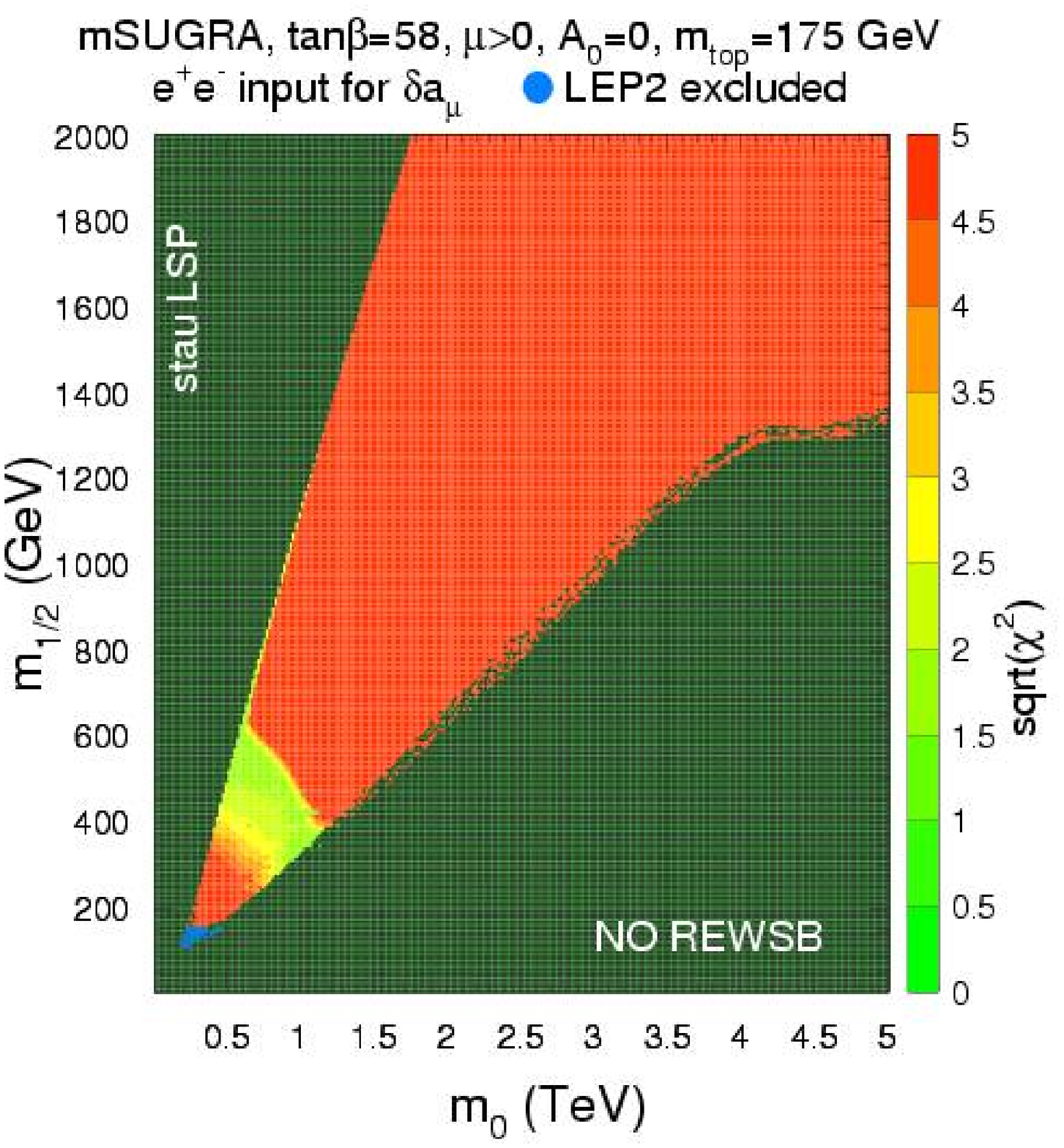,width=7.5cm}
\hspace*{-0.5cm}
\caption{Plot of regions of $\sqrt{\chi^2}$ in the mSUGRA model for
$A_0=0$, $\mu >0$, and $\tan\beta =10,\ 30,\ 55$ and 58.
The green regions have low $\chi^2$, while red regions have
high $\chi^2$. Yellow is intermediate.
}
\label{fig:msugra}
}
\FIGURE{%
\hspace*{-0.5cm}
\epsfig{file=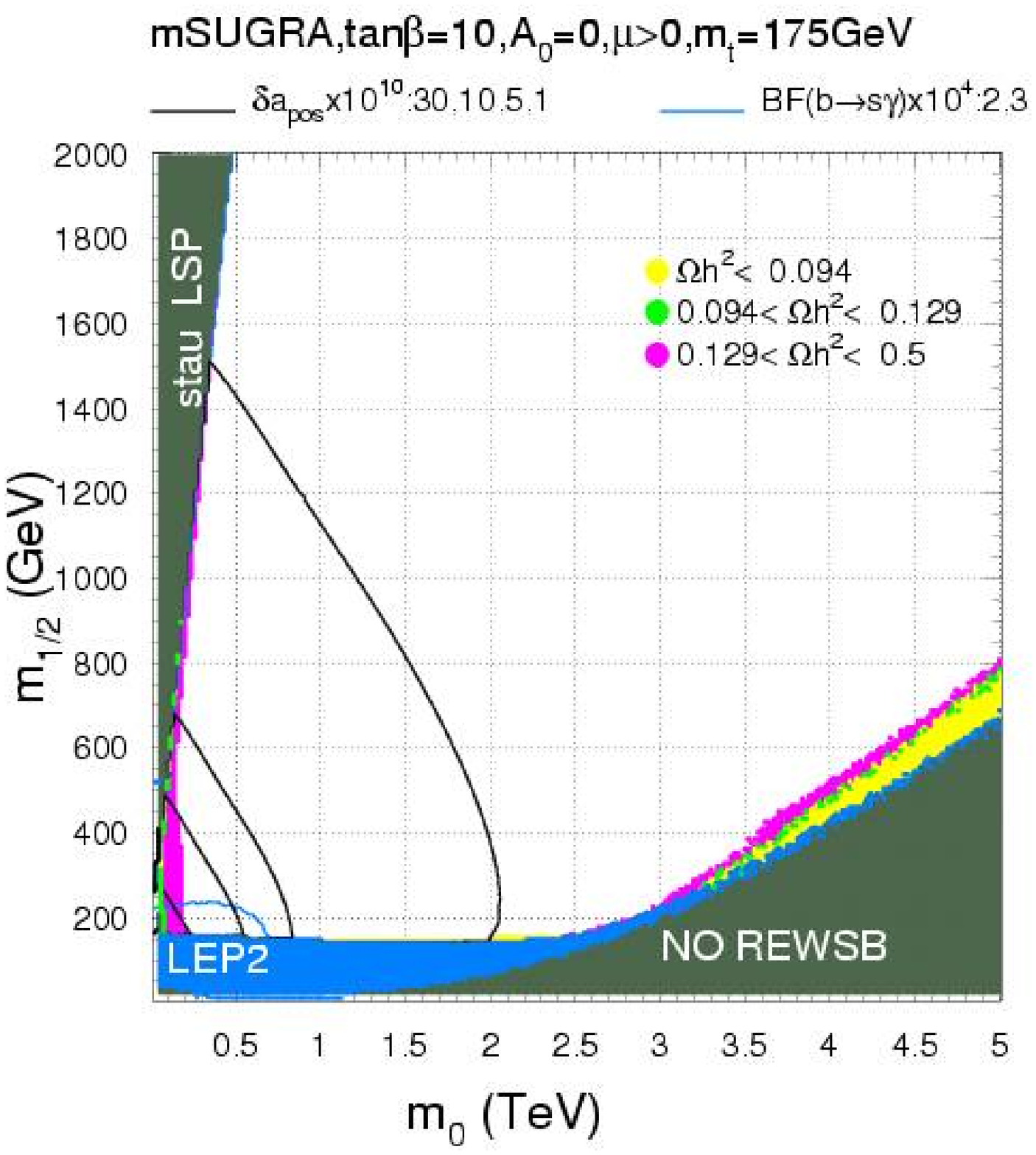,width=7.5cm}
\hspace*{-0.5cm}
\epsfig{file=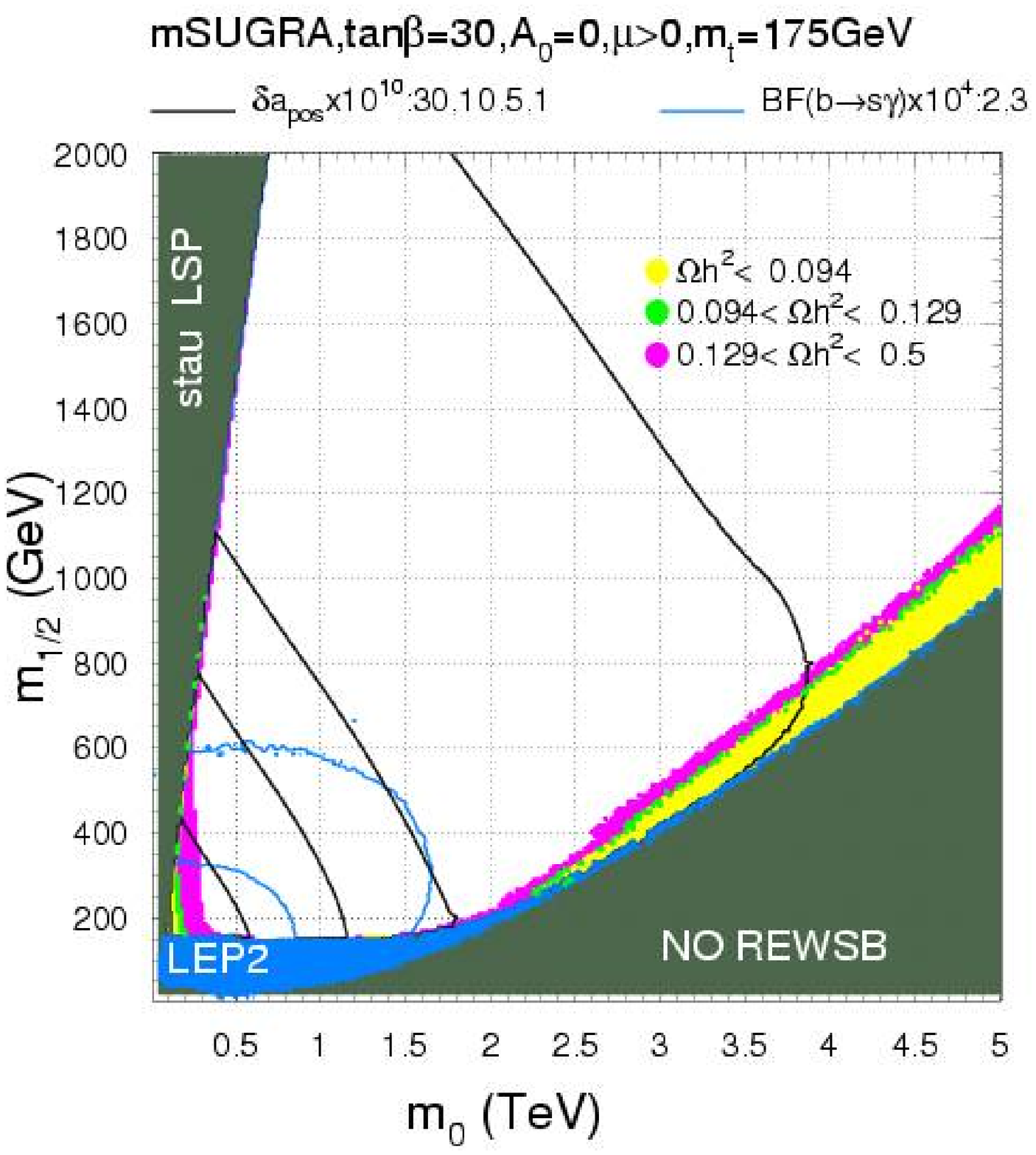,width=7.5cm}
\hspace*{-0.5cm}
\\
\hspace*{-0.5cm}
\epsfig{file=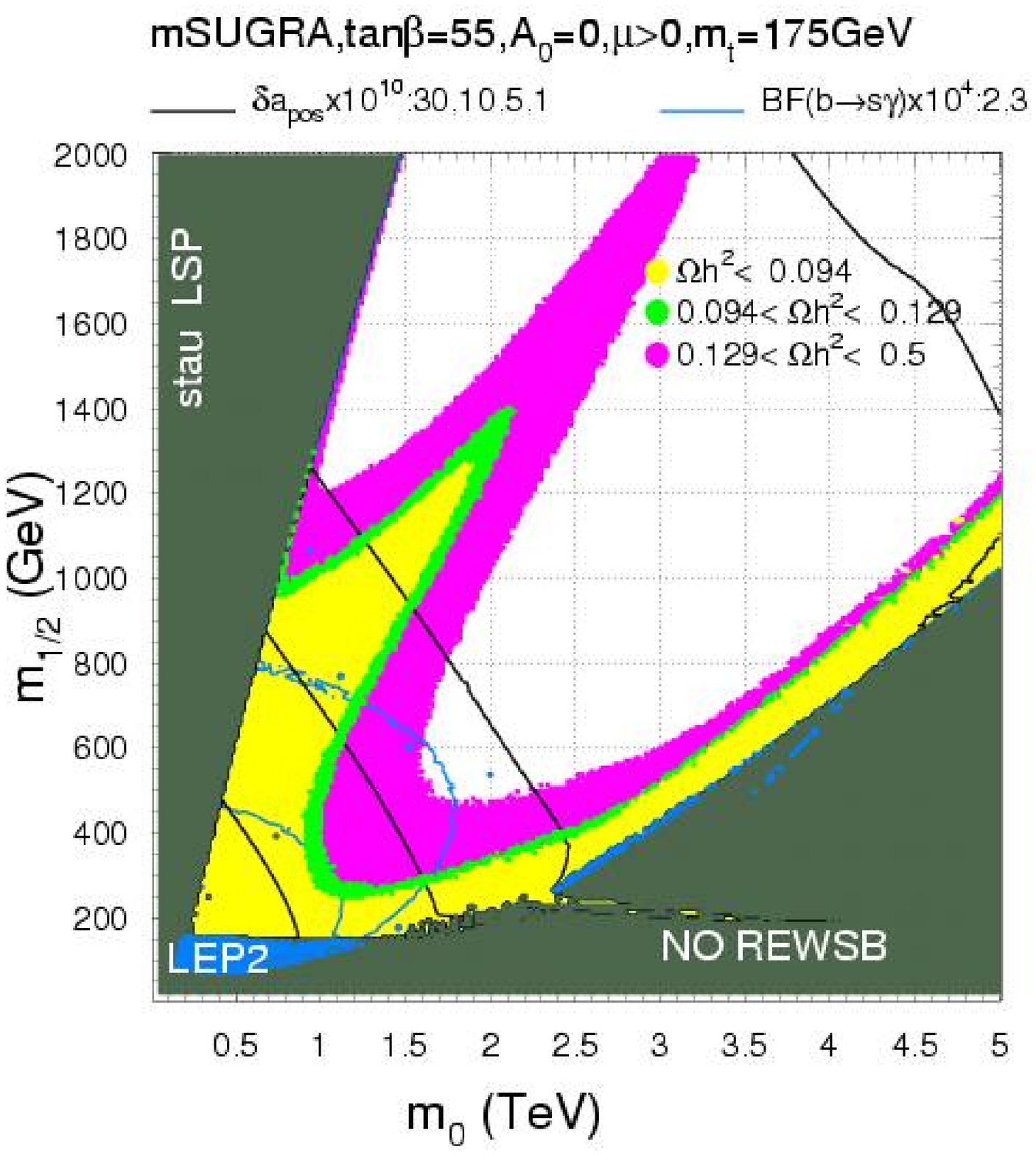,width=7.5cm}
\hspace*{-0.5cm}
\epsfig{file=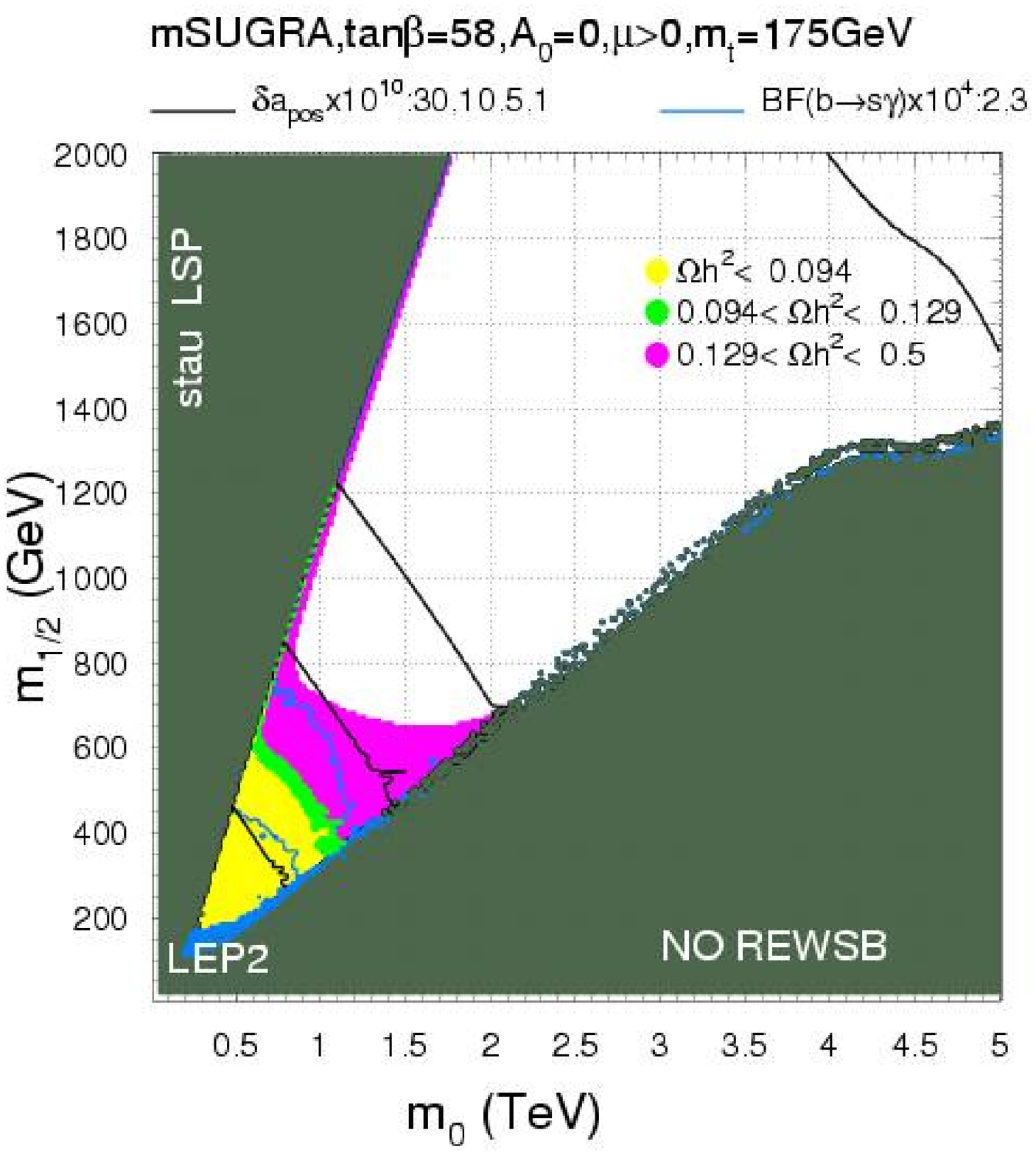,width=7.5cm}
\hspace*{-0.5cm}
\caption{Contour levels for $BF(b\to s\gamma)$, $a_\mu$ and $\Omega h^2$
in the mSUGRA model for
$A_0=0$, $\mu >0$, and $\tan\beta =10,\ 30,\ 55$ and 58.
Magenta region excluded by LEP2 searches.
}
\label{fig:msugrac}
}

In the $\tan\beta =10$ case, it can be seen that almost all the 
$m_0\ vs.\ m_{1/2}$ plane has very large $\chi^2$. This arises because 
in general an overabundance of dark matter is produced in the 
early universe, 
and the relic density $\Omega_{\tz_1}h^2$ is beyond WMAP limits.
There is a very narrow sliver of yellow at $m_{1/2}\sim 150$ GeV
(just beyond the LEP2 limit) where $2m_{\tz_1}\simeq m_h$, and neutralinos
can annihilate through the narrow light Higgs resonance. 
In addition, there is an orange/yellow region at high $m_0$ 
at the edge of parameter space (the HB/FP region), with an intermediate
value of $\chi^2$. In an earlier study\cite{sug_chi2}, 
this region was found to have a low
$\chi^2$ value. In this case, however, the $3\sigma$ deviation from the SM of
$a_\mu$ tends to disfavor the HB/FP region. In the HB/FP region, 
sleptons are so heavy (typically 3-5 TeV), that SUSY contributions to 
$a_\mu$ are tiny, and the prediction is that $a_\mu$ should be in near 
accord with the SM calculation. The remaining green region is the narrow sliver
that constitutes the stau co-annihilation region, 
barely visible at the left hand edge of parameter space 
adjacent to where $\ttau_1$ becomes the LSP. The overall situation is similar
when we move to $\tan\beta =30$: the only low $\chi^2$ region is 
in the stau co-annihilation corridor. 

Once we move to very large $\tan\beta$ values,
as shown in the third frame, then the $A$-annihilation funnel
becomes visible, and some large regions of moderately low $\chi^2$ 
appear around $m_0,\ m_{1/2}\sim 500,\ 600$ GeV and also at
$1500,\ 200$ GeV. While the $A$-annihilation funnel extends
over a broad region of parameter space, the upper and lower ends of
the funnel are disfavored: basically, if sparticles become too 
heavy (the upper end), then
$\Delta a_\mu$ becomes too small, 
while if sparticles become too light (the lower end), 
then $BF (b\to s\gamma )$
deviates too much from its central value. The final frame shows results for
$\tan\beta =58$, close to the point where parameter space begins to collapse
due to an inappropriate breakdown of electroweak symmetry. 
The relic density is low over much of the parameter space 
since $2m_{\tz_1}\sim m_A$, so a region of relatively low $\chi^2$ appears
at modest $m_0$ and $m_{1/2}$ values.

Our conclusion for the mSUGRA model is that almost all of
parameter space is excluded or at least disfavored by the combination
of the WMAP $\Omega_{\tz_1}h^2$ limit, the new $\Delta a_\mu$
value, and the $BF(b\to s\gamma )$ value. The $\Omega_{\tz_1}h^2$ 
constraint only allows the several regions of parameter space mentioned 
in Sec. \ref{sec:intro}, while $BF(b\to s\gamma )$ favors large
third generation squark masses to suppress 
SUSY contributions to $b\to s\gamma $
decay, and $\Delta a_\mu$ favors relatively light second generation
slepton masses, to give a significant deviation of $(g-2)_\mu$ from
the SM value. The only surviving regions with relatively low
$\sqrt{\chi^2}\alt 2$ are the stau co-annihilation region, and 
intermediate portions of the $A$-annihilation funnel at very large
values of $\tan\beta$. The previously favored HB/FP region from
Ref. \cite{sug_chi2} is now disfavored because it is unable to generate
a significant deviation of $(g-2)_\mu$ from the SM prediction
which uses the $e^+e^-\to hadrons$ data for 
evaluating the hadronic vacuum polarization graphs. 
(We note that should the hadronic vacuum polarization determination 
using $\tau$ decay data turn out to be correct, then the
HB/FP region will appear in a more favorable light!)

\section{$\chi^2$ fit of NMH SUGRA model}
\label{sec:nmh}

It is easy to see that $\Delta a_\mu$ favors light second generation
sleptons, while $BF (b\to s\gamma )$ prefers heavy third generation squarks.
This situation is hard to realize in the mSUGRA model, and may be an
indication that one must move beyond the assumption of 
universality, wherein each generation has a common mass at
$Q=M_{GUT}$. In this section, we advocate expanding the parameter
set of the mSUGRA model to the following values:
\be
m_0(1),\ m_0(3),\ m_H,\ m_{1/2},\ A_0,\ \tan\beta,\ sign (\mu ) .
\label{eq:nups}
\ee
In the above, $m_0(1)$ will be a common scalar mass of all {\it first}
generation scalars at $Q=M_{GUT}$, while $m_0(3)$ is the common
mass of all {\it third} generation scalars at $M_{GUT}$. $m_H$ is the 
common Higgs mass at $M_{GUT}$. The above parameter set is well motivated
in $SO(10)$ SUSY GUT models, where the two MSSM Higgs doublets
typically occupy a {\bf 10} of $SO(10)$, and each generation
of scalars, along with a SM gauge singlet $N$ occupies the 
{\bf 16} dimensional spinorial representation of $SO(10)$.
Thus, the $SO(10)$ gauge symmetry enforces universality within 
each generation, although each generation may be split one from
another even when $SO(10)$ is unbroken. In fact, it is well known 
that in models where $SO(10)$ is valid beyond the GUT scale, 
at least the third generation of scalars will evolve to independent
mass values even if all generations begin with a common mass 
somewhere above $M_{GUT}$; see for instance, Ref. \cite{bdqt}.

The step of breaking generational universality must be 
taken with some caution, 
since in general it can lead to violations of constraints 
from FCNC processes.
The most stringent of the FCNC limits comes from contributions to
the $K_L-K_S$ mass difference, and from limits on $\mu\to e\gamma$
decay. These bounds, however, apply to splittings between the first and
second generations. 

Constraints from FCNC processes are usually presented as bounds on
off-diagonal terms of squark or slepton soft SUSY breaking(SSB)
mass matrices or trilinear 
soft breaking terms in the super-CKM basis\cite{masiero,hagelin,Misiak:1997ei}.
Here, we are interested in a ``proof-of-principle'' that large splittings 
between first and third generation scalars can exist and satisfy
constraints from FCNC processes. 
We will assume a scenario of ``minimal mixing'', where
SSB mass-squared and trilinear matrices are diagonal at 
the electroweak scale, 
and that off-diagonal elements are generated {\it only } by rotation
of the weak scale SSB matrices to the super-CKM basis\footnote{Arbitrarily 
large off-diagonal
SSB elements would lead to fatal conflicts with FCNC bounds.}. 
In addition, in our constraint calculations we neglect
the contribution of trilinear soft terms, since these contributions are often
smaller than those from the mass matrices. 
An explicit case study is shown in the Appendix.

Within the minimal mixing scenario,
the limits from the $K_L-K_S=\Delta m_K$ mass difference in the super-CKM basis
can be translated to limits on mass splittings of diagonal 
soft SUSY breaking mass terms. In the case of equal squark and gluino masses,
these bounds are roughly\cite{Misiak:1997ei}
\be
|m_{\tilde q}(1)-m_{\tilde q}(2)|\alt 2m_c \frac{m_{\tq}^2}{M_W^2},
\ee
which yields mass splittings for weak scale 
SSB squark masses of order the
charm quark mass ($m_c$)
if the average squark mass is
$m_{\tq}\simeq M_W$.
If squark masses are much heavier, 
the limits become less stringent. Allowed regions of the 
$m_{\tq}(1)\ vs.\ m_{\tq}(2)$ plane are shown in Fig. \ref{fig:kkmix}
for several different gluino masses. The limits are obtained from 
the complete gluino-squark box diagram computation, where it is 
conservatively required
that $|\Delta m_K^{SUSY}|\le \Delta m_K^{exp}$, and $\Delta m_K^{exp}$
is the measured $K_L-K_S$ mass difference. The limits are presented
assuming {\it i}.) just a contribution from left squark-left squark
mass terms (LL: green-shaded region), or more stringently {\it ii}.) from
the presence of both left-left and right-right (RR) squark mass terms
(blue-shaded region).
The less stringent constraint assumes
that there is splitting only between soft SUSY breaking masses for left
squarks of the 1st and 2nd generations, while the corresponding right
squarks remain degenerate. In this way, non-degeneracy of up to 50\% can be
allowed for squarks in the TeV range. 
It can be seen from Fig. \ref{fig:kkmix}
that a liberal allowance of non-degeneracy does not work with either
scenario if squarks $\sim 100$ GeV, 
because both
areas converge to a line of nearly perfect degeneracy, which was
the main motivation for the mSUGRA model.
In this paper, in order to accomodate the kaon mass difference constraint, 
we will, without further discussion, maintain
universality, but just between the first and second generations:
$m_0(1)\simeq m_0(2)$.
\FIGURE{\epsfig{file=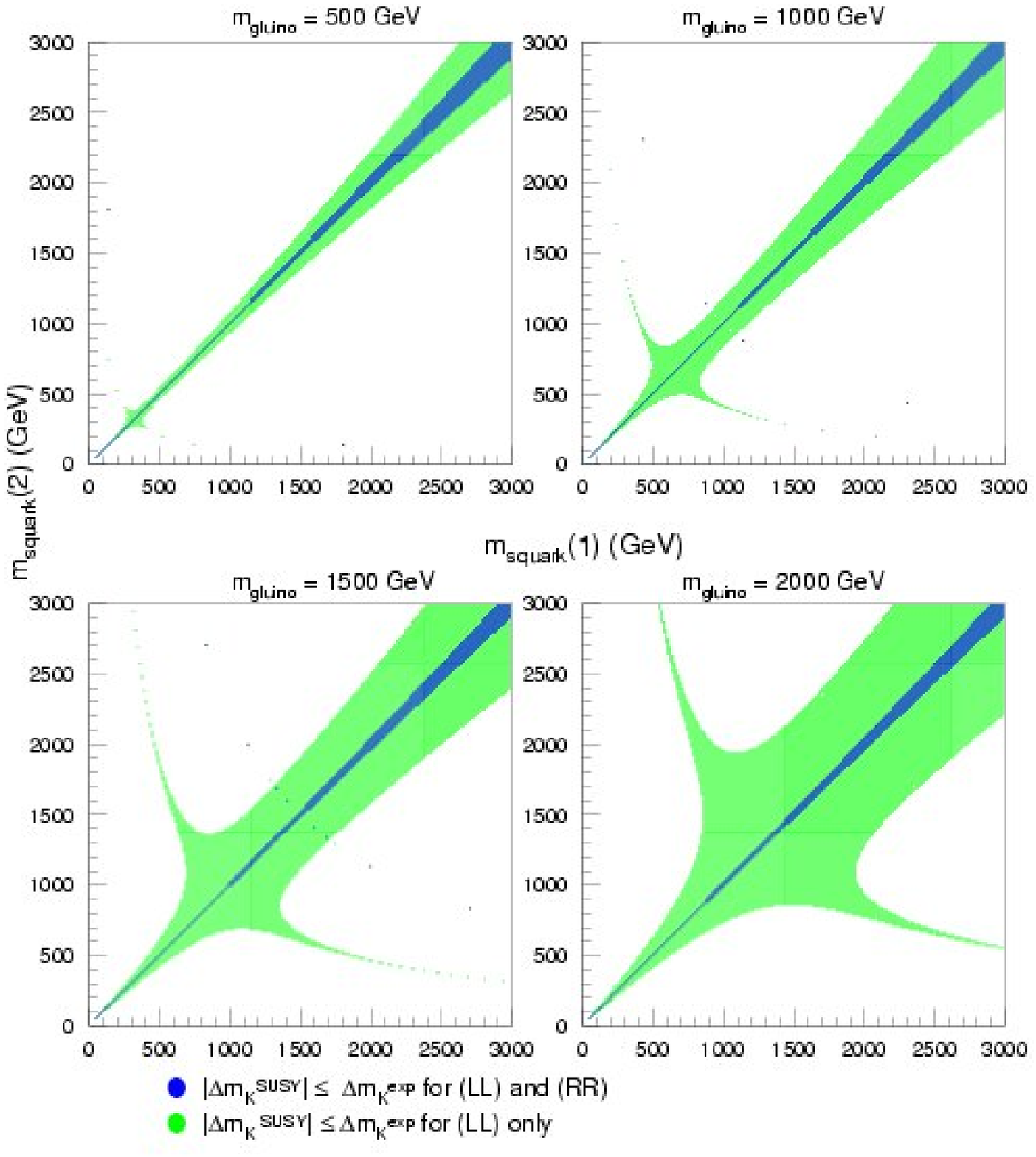,width=12cm}
\caption{Constraints on first and second generation squark masses from
$\Delta m_{K}$.
}
\label{fig:kkmix}
}

Splitting the third generation from the first and second can also
potentially lead  to violations of FCNC processes.  One of the main
experimentally measured bounds on FCNC processes in this case comes from
$B^0_H-\bar{B^0_L}$ mass splitting. 
As in the kaon case, we neglect all
but the largest contribution to $\Delta m_B$
coming from gluino-squark box diagrams. In this case, for
equal gluino and squark masses, the soft term mass splitting limit is
\be
|m_{\tq}(1)-m_{\tq}(3)|\alt \frac{m_{\tq}^2}{M_W},
\ee
which is much less restrictive than the kaon case, for both low and high
squark masses.
We adopt the result for gluino box diagrams from Ref.~\cite{hagelin} to
calculate the SUSY contribution to $\Delta m_{B_d}$. The result for
various values of $m_{\tg}$ is presented in the $m_{\tq}(1)\ vs.\
m_{\tq}(3)$ plane  in  Fig.~\ref{fig:bbmix}, where $m_{\tq}(i)$ is the
weak scale squark mass for generation $i$.
The magenta regions give values of 
$|\Delta m_B^{SUSY}|<\Delta m_B^{exp}/1000$, while green gives
$|\Delta m_B^{SUSY}|<\Delta m_B^{exp}/100$, and yellow gives
$|\Delta m_B^{SUSY}|<\Delta m_B^{exp}/10$.
From the figure, it is clear that practically the whole  parameter space
of NMH SUGRA model is allowed by the $\Delta m_{B_d}$ constraint.
\FIGURE{\epsfig{%
file=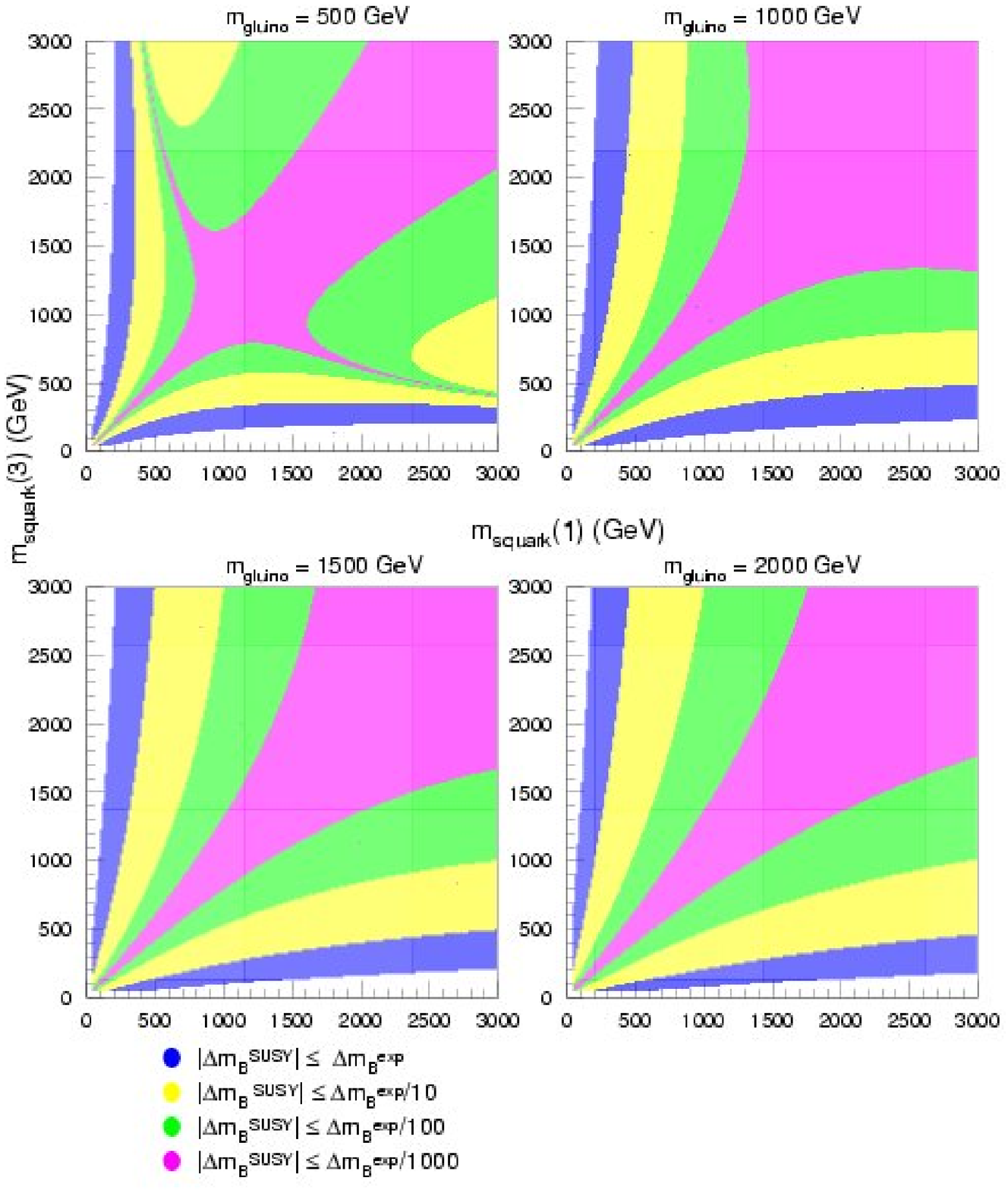,width=12cm}
\caption{Constraints on first and third generation squark masses from
$\Delta m_{B_d}$.
}
\label{fig:bbmix}
}

We have also checked the potential constraint on $m_0(2)-m_0(3)$
splitting from $BF(b\to s\gamma)$ measurements. These constraints
also yield weak bounds on generational splitting; see the Appendix for
an explicit calculation. 
More details can be found, for example, in 
Ref.~\cite{Misiak:1997ei} (see Eqs. (53-54)) 
and Ref.~\cite{fcnc} (see Eq. (19)), along with discussion of
these bounds.

Independent of which precise limit on $\Delta m_B$ we choose, it is 
easy to see from Fig.~\ref{fig:bbmix} 
that any constraint on degeneracy of weak scale squark masses
applies only if squark masses are quite light, well below the TeV scale.
As we move to TeV level squark (and gluino) masses and beyond, 
the limits become essentially non-existent. This corresponds to a partial
decoupling solution to the SUSY flavor problem. 

However, 
even if squark mass splittings are very large at the scale 
$Q=M_{GUT}$, the weak scale mass splittings are often much smaller.
As an example, we
illustrate the running of soft SUSY breaking parameters in 
Fig. \ref{fig:masses}, where we choose
\begin{eqnarray}
&& m_0(1)=100\ {\rm GeV}, \ \ m_0(3)=m_H=1400\ {\rm GeV}, \ \ 
m_{1/2}=550\ {\rm GeV},
\nonumber\\
&& A_0=0,\ \ \tan\beta =30, \ \ \mu >0, \ \ m_t=175\ {\rm GeV}.
\label{eq:sample}
\end{eqnarray}
While first generation slepton masses are renormalized
to values of $m_{\tell}\sim 100-300$ GeV, the first generation
squark masses evolve to weak scales values in the TeV range, since the 
QCD contributions to RG running are large~\footnote{A simple
approximate
formula relating weak scale to GUT scale squark and slepton masses is that
$m_{\tq}^2\simeq m_0^2+(5-6)m_{1/2}^2$, while 
$m_{\tell}^2\simeq m_0^2+(0.15-0.5)m_{1/2}^2$.}. 
Meanwhile, the running of third generation squarks is actually
suppressed somewhat, owing to the large top quark Yukawa coupling.
The result is that, even beginning with a large splitting in generations
at $Q=M_{GUT}$, a much less severe splitting amongst squarks 
may be obtained at the weak scale.
Of course, the splitting amongst sleptons remains rather large, and could give
large contributions to $\tau\to e\gamma$ and $\tau\to \mu\gamma$ decay.
However, the current experimental limits from flavor changing 
radiative $\tau$ decay are not overly constraining\cite{masiero}.

\FIGURE{\epsfig{file=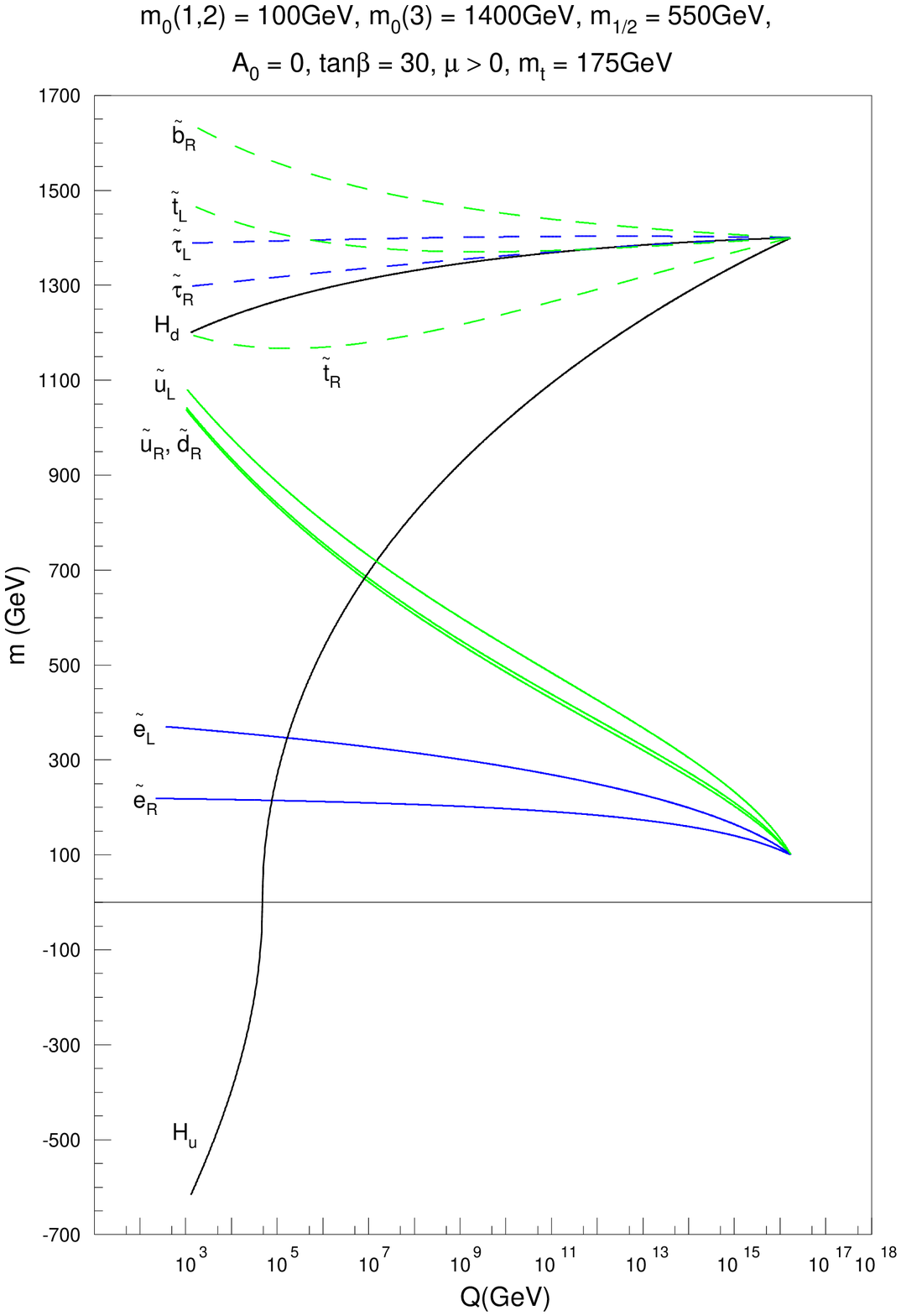,width=10cm}
\vspace*{0.3cm}
\caption{Evolution of soft SUSY breaking masses in the NMH SUGRA
model. 
}
\label{fig:masses}
}

Motivated by above considerations, we next generate random points in the
parameter space given in Eq. \ref{eq:nups}, and again calculate
the resultant $\chi^2$ values. 
We have scanned over the following region of parameter space for $sign(\mu)>0$:
\begin{eqnarray}
&&m_0(1): 0-\ 3\ \mbox{TeV}, \ \  m_0(3): 0-\ 3\ \mbox{TeV}, \ \ 
m_{H}: 0-\ 3\ \mbox{TeV}, 
\nonumber\\
&&m_{1/2}: 0-\ 1\ \mbox{TeV}, \ \ A_0:   -3-\ 3\ \mbox{TeV}, \ \ 
\tan\beta : 2-\ 60 .
\end{eqnarray}
The colors of the points correspond
to those of Fig. \ref{fig:msugra}, so that green gives low $\chi^2$, 
red gives high $\chi^2$, and yellow gives intermediate values.
A variety of frames showing correlations amongst the parameters are
shown in Fig. \ref{fig:scan}. 
Only points satisfying the LEP2 constraints have been plotted.
This is why, for example,
$m_{1/2}$ is bounded from below at 
$\sim 200$GeV, corresponding to the limit on the 
chargino mass of $m_{\tw_1}>103.5$ GeV.

From the top row of figures, we see that rather low values of 
$m_0(1)\sim 0-400$ GeV are preferred. These parameter values 
give sufficiently light
smuon and mu sneutrino masses so as to fulfill the $(g-2)_\mu$
constraint. Values of $m_{1/2}\sim 400-800$ GeV are preferred, although
little preference is shown for $A_0$. 
One can also see little preference for $\tan\beta$ so long as 
$m_0(1)$ remains small. For larger values of $m_0(1)$, there is a preference
for very high values of $\tan\beta\sim 50-60$, as in the mSUGRA model case.

The plots of the second row show that there is some preference for
$m_H\sim m_0(3)$, and that there is a preference for $m_0(1)\ll m_0(3)$.
The third row shows that while $m_0(1)\sim 0-400$ GeV, 
$m_0(3)\sim 500-3000$ GeV is preferred. Further, the last frame of this row
shows that while low $\chi^2$ points can occur at any $\tan\beta$
value if there is a large $m_0(1)-m_0(3)$ mass splitting, that mainly
large $\tan\beta$ is preferred if the generational mass splitting
is small (which takes us back towards the mSUGRA case).
If may also be pointed out that SUSY IMH models
are greatly disfavored.
Finally, the last row of plots shows the mechanism for annihilating
neutralinos in the early universe. The first frame shows a region of
low $\chi^2$ where $2m_{\tz_1}\simeq m_A$, {\it i.e.} the $A$-funnel at
large $\tan\beta$, which occurs in the subset of mSUGRA like models.
The second frame shows that some models get rid of neutralinos by 
stau co-annihilation, while the third frame shows that now many models
may also destroy neutralinos in the early universe
via $\tz_1-\te$ and
$\tz_1 -\tmu$ co-annihilation.
\FIGURE{\epsfig{file=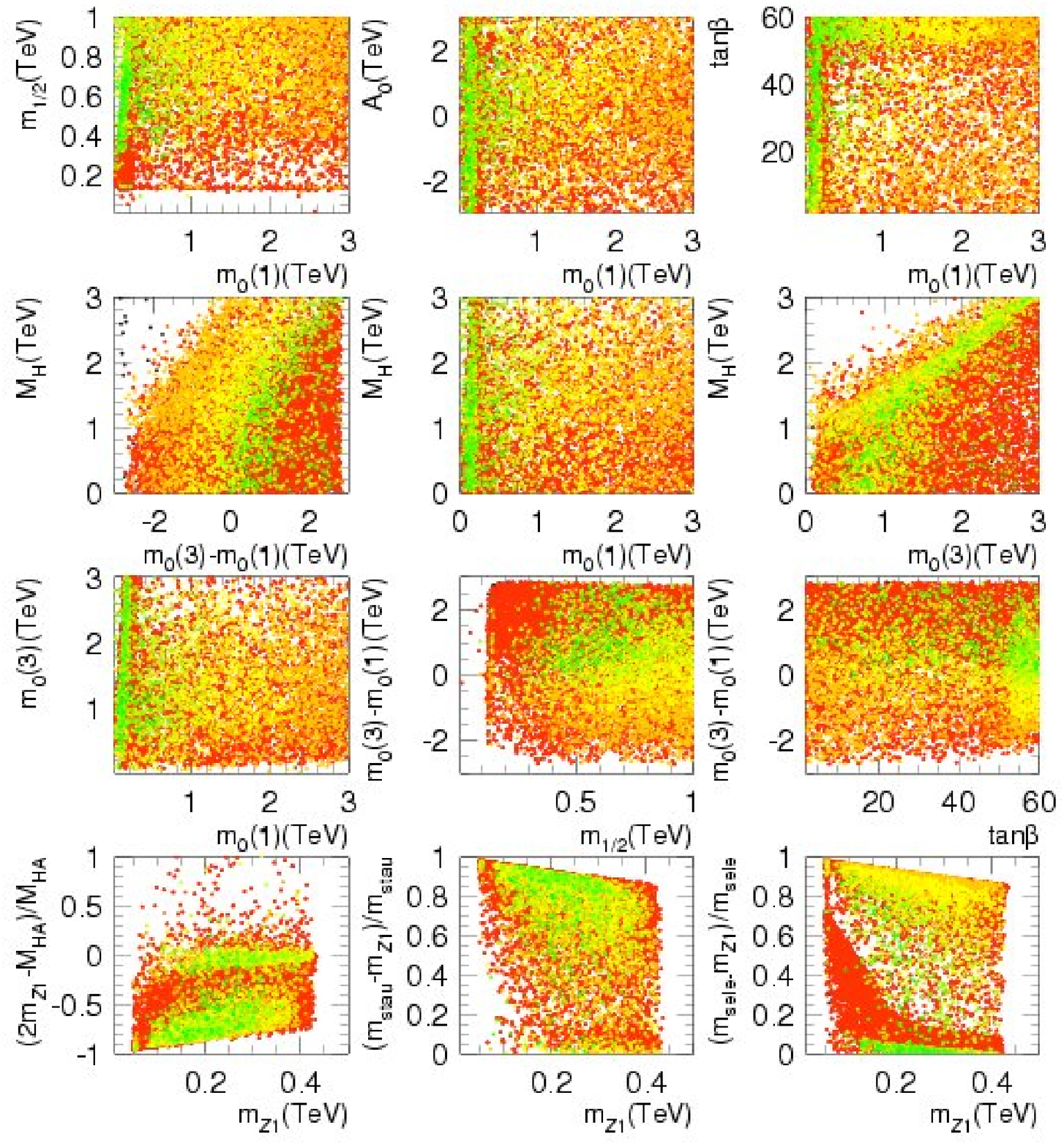,width=15cm}
\vspace*{-0.8cm}
\caption{Scan over parameter space of NMH SUGRA model. 
Green points have low $\chi^2$, while red points have high $\chi^2$;
yellow points have itermediate $\chi^2$.
}
\label{fig:scan}
}

Motivated by the above scan, we next adopt the value $m_H= m_0(3)$
(to reduce parameter freedom), and plot the $\sqrt{\chi^2}$ values
in the $m_0(3)\ vs.\ m_{1/2}$ plane in Fig.~\ref{fig:nmh_10}, 
for $m_0(1)$ values of
50, 100 and 200GeV, with $\tan\beta =10$, $A_0=0$ and $\mu >0$.
The corresponding contour plots of
$BF(b\to s\gamma)$, $a_\mu$ and $\Omega h^2$ are shown in 
Fig. \ref{fig:nmh_10c}.

The most striking feature of the plots is that most of the area displayed is
excluded. In this case, slepton masses are quite light, in the vicinity
of a few hundred GeV. As $m_{1/2}$ increases, ultimately $m_{\tz_1}$
becomes greater than $m_{\tell}$, and one violates the cosmological constraint
on stable charged relics from the Big Bang. Of the surviving parameter space, 
we find significant regions with relatively low $\chi^2$. The plot with
$m_0(1)=50$ GeV has a rather broad band of low $\chi^2$. In this case,
neutralinos in the early universe can annihilate by a combination
of $t$-channel slepton exchange (as in the bulk region of mSUGRA), and
by neutralino-slepton co-annihilation. In addition, smuons and mu sneutrinos 
are relatively light, giving a large, positive contribution to
$\Delta a_\mu$, while top squarks inhabit the TeV and beyond range, 
effectively suppressing anomalous contributions to $BF(b\to s\gamma )$.
In addition to the region with low values of $m_0(3)$ and low $m_{1/2}$,
a thin strip of the HB/FP region survives, where the small value
of $\mu$ that is generated leads to a Higgsino-like LSP. In this case, 
the HB/FP has moderate but not low $\chi^2$ values because the contribution
to $\Delta a_\mu$ is low. Finally, we note in this plot that the 
stau co-annihilation region is somewhat enhanced as well on the left most
boundary of parameter space. 
This is because neutralinos can annihilate via a combination 
of $t$-channel slepton exchange, and also stau co-annihilation.

\FIGURE{%
\epsfig{file=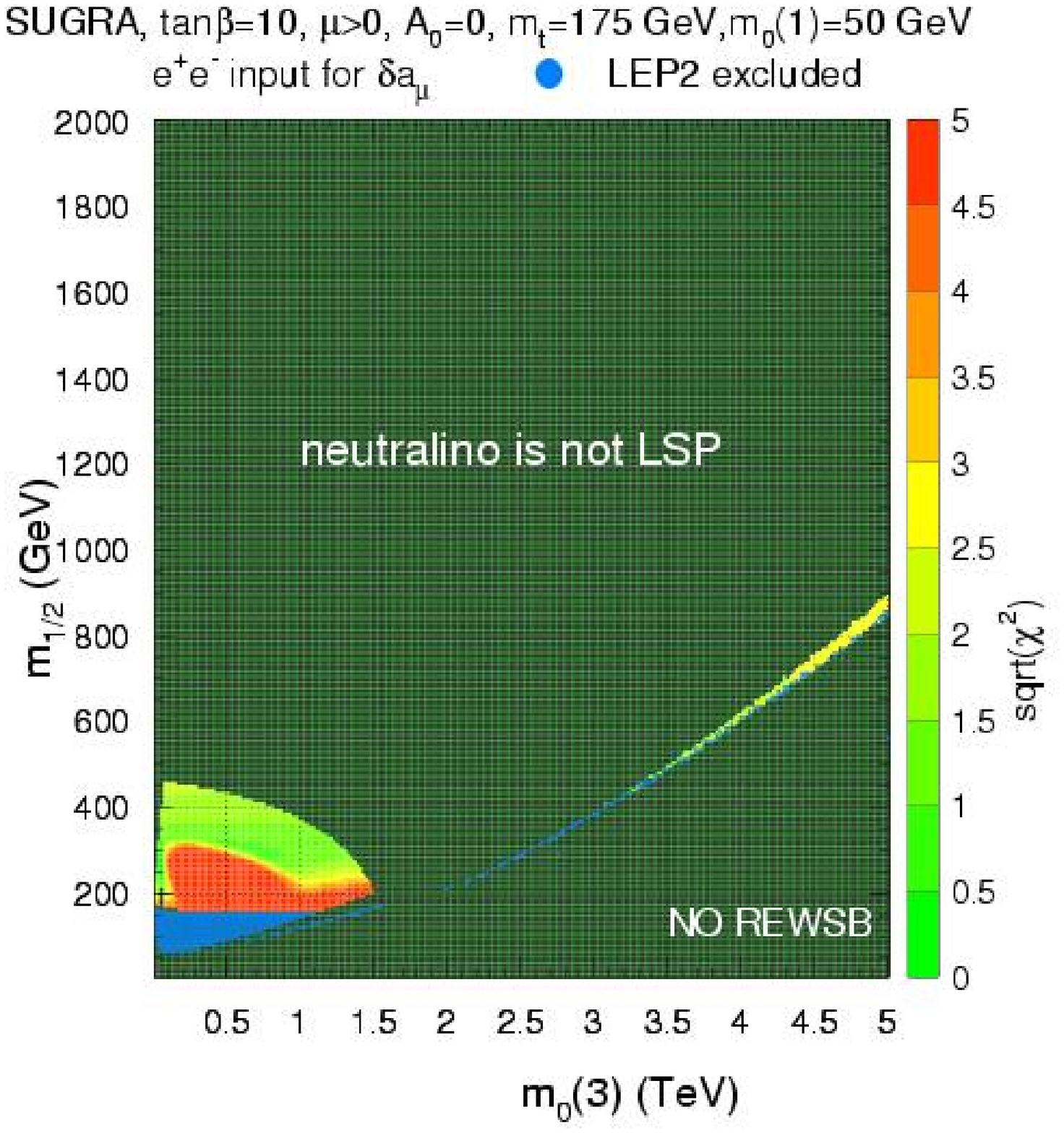,width=7cm}
\epsfig{file=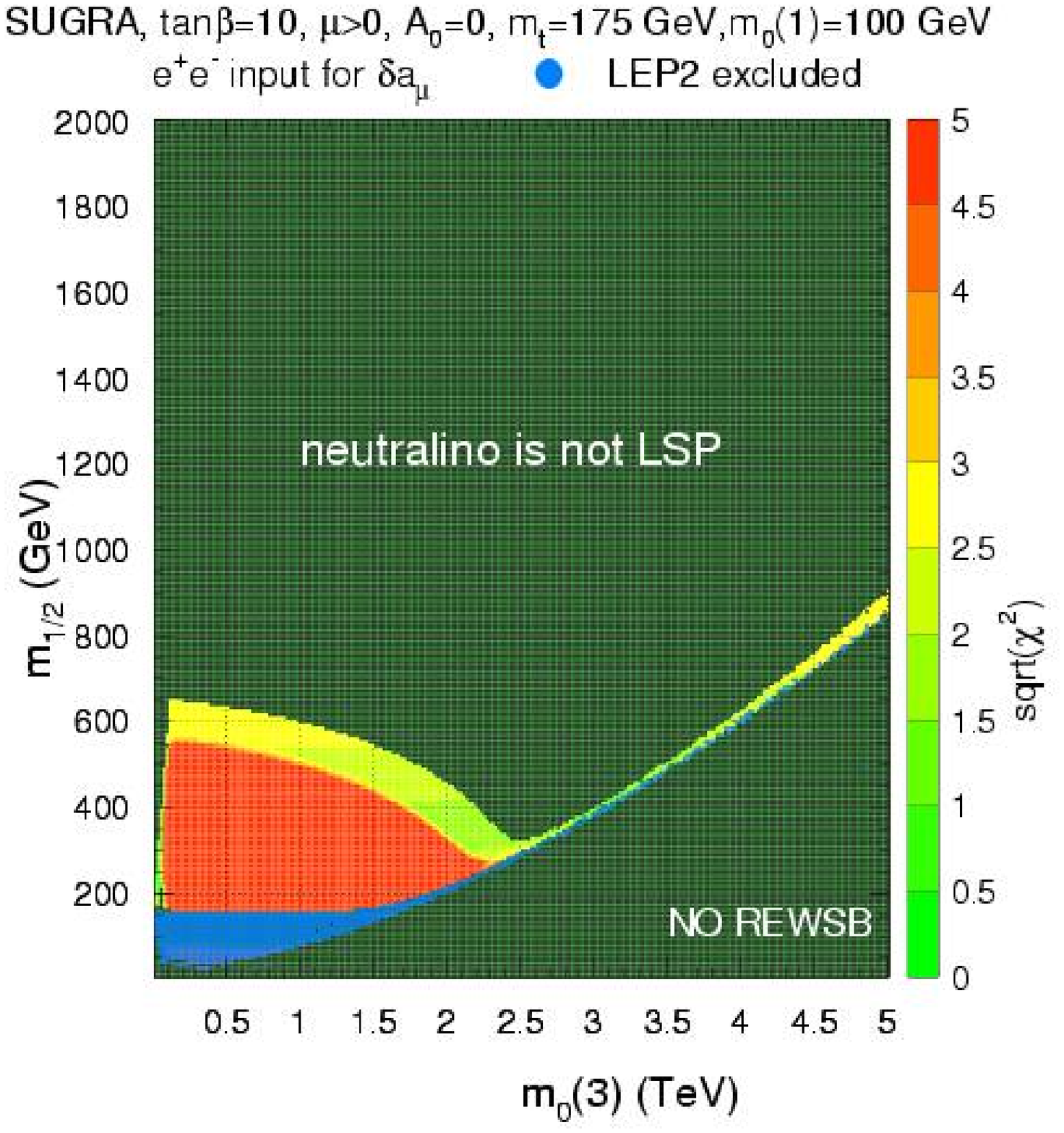,width=7cm}\\
\epsfig{file=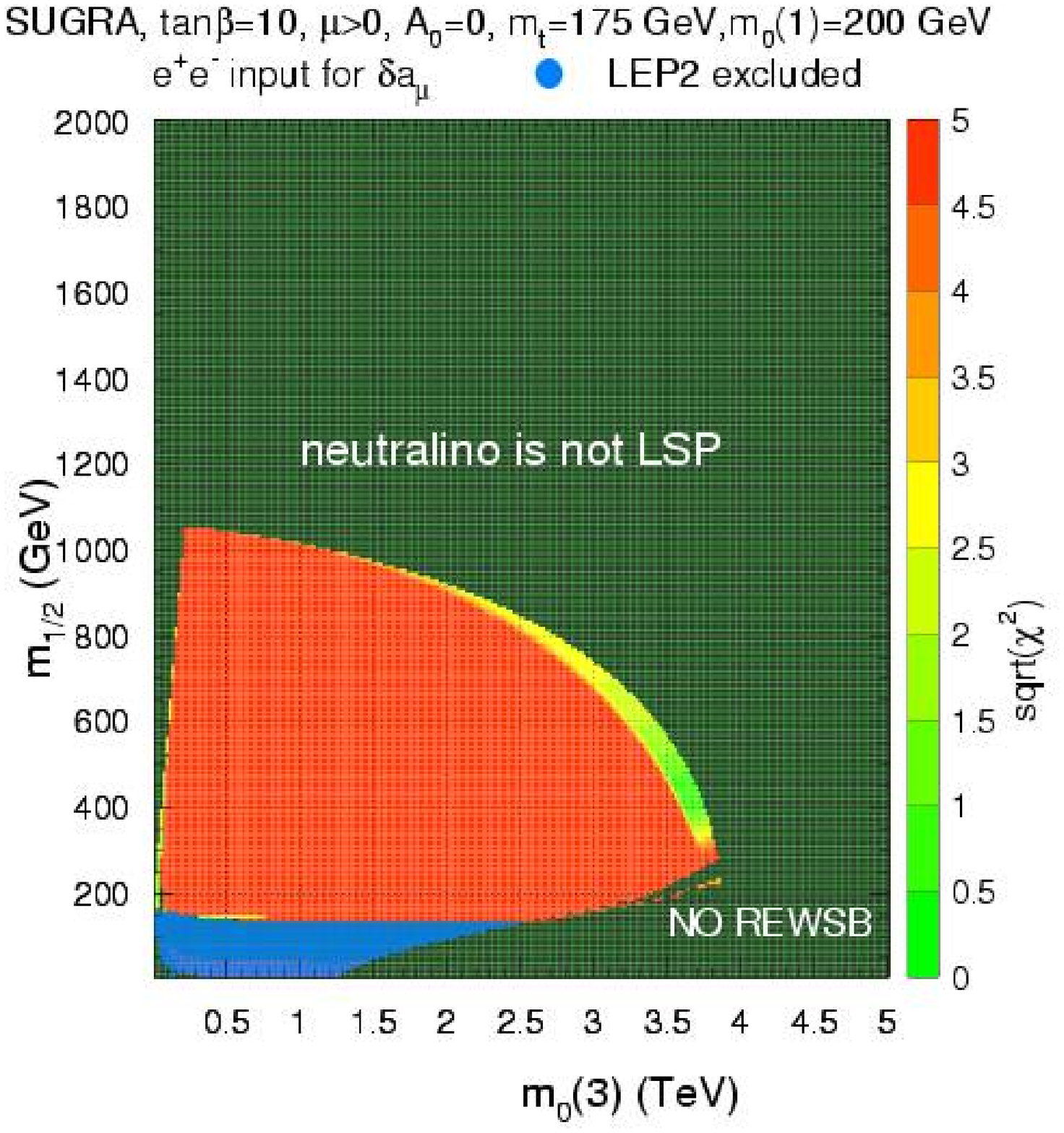,width=7cm}
\caption{Plot of regions of $\sqrt{\chi^2}$ in the 
$m_0(3)\ vs. m_{1/2}$ plane of the NMH SUGRA model for
$m_0(1,2)=50$, 100 and 200 GeV with 
$A_0=0$, $\mu >0$, and $\tan\beta =10$.
The green regions have low $\chi^2$, while red regions have
high $\chi^2$. Yellow is intermediate.
}
\label{fig:nmh_10}
}

\FIGURE{%
\epsfig{file=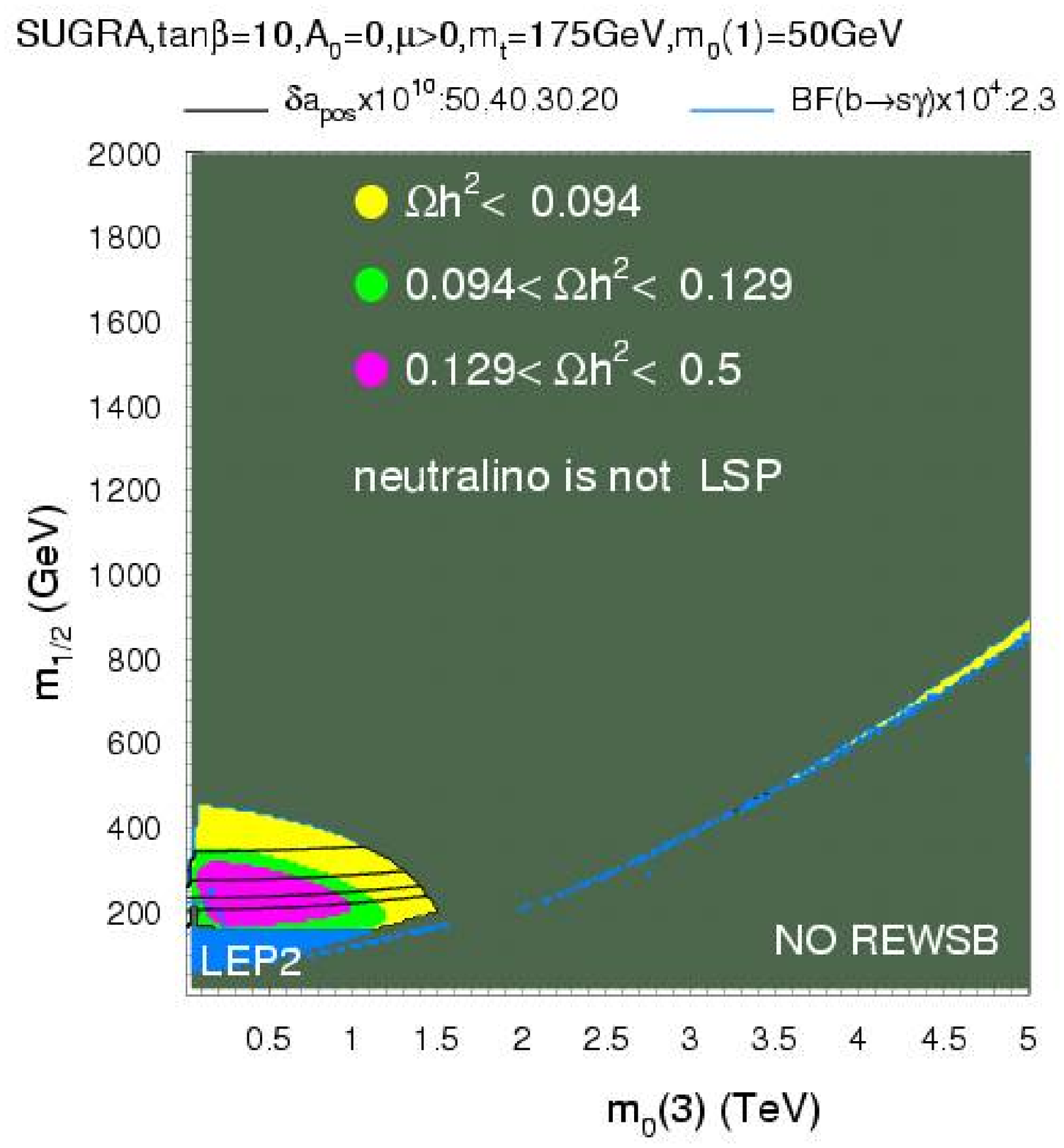,width=7cm}
\epsfig{file=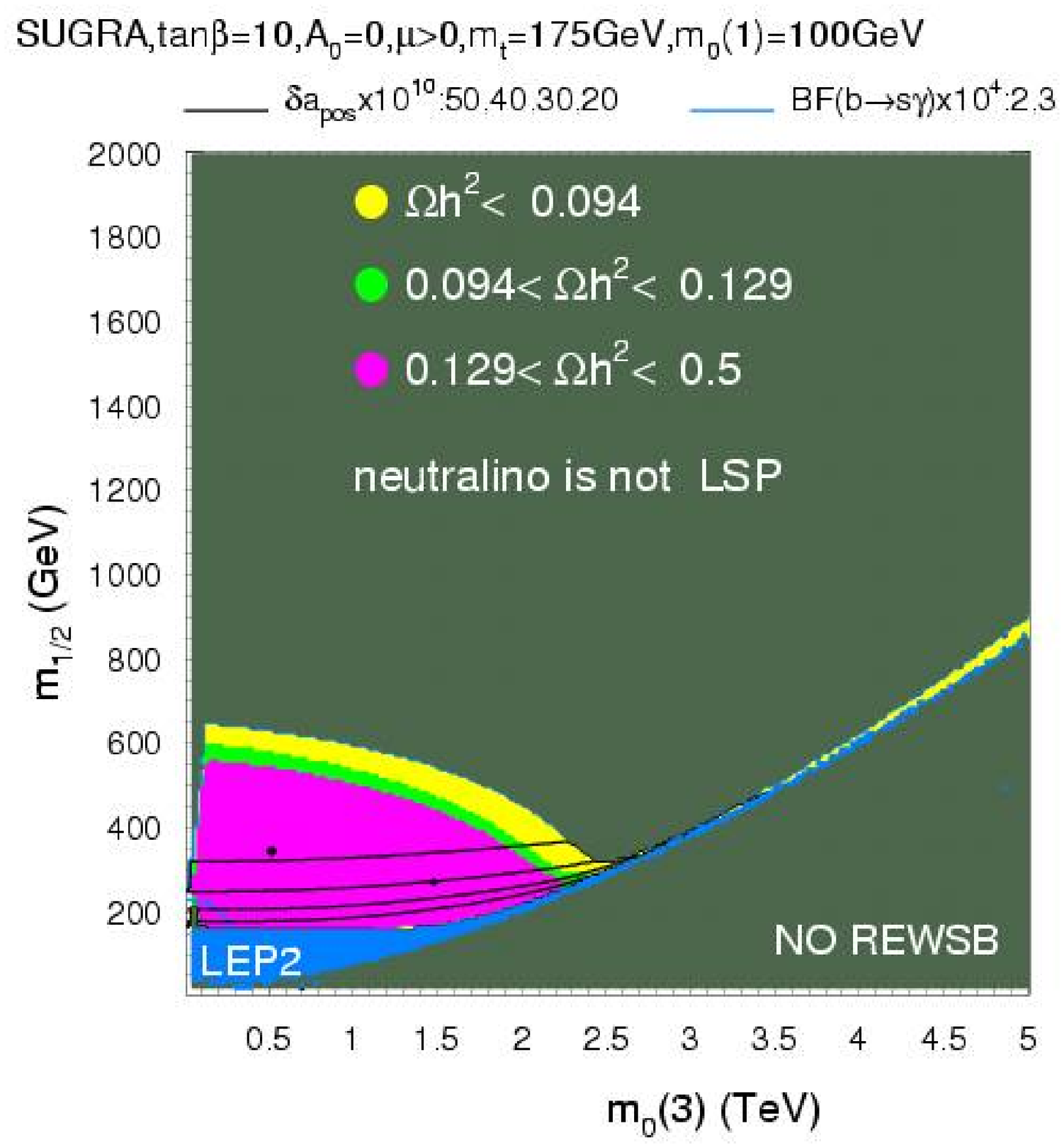,width=7cm}\\
\epsfig{file=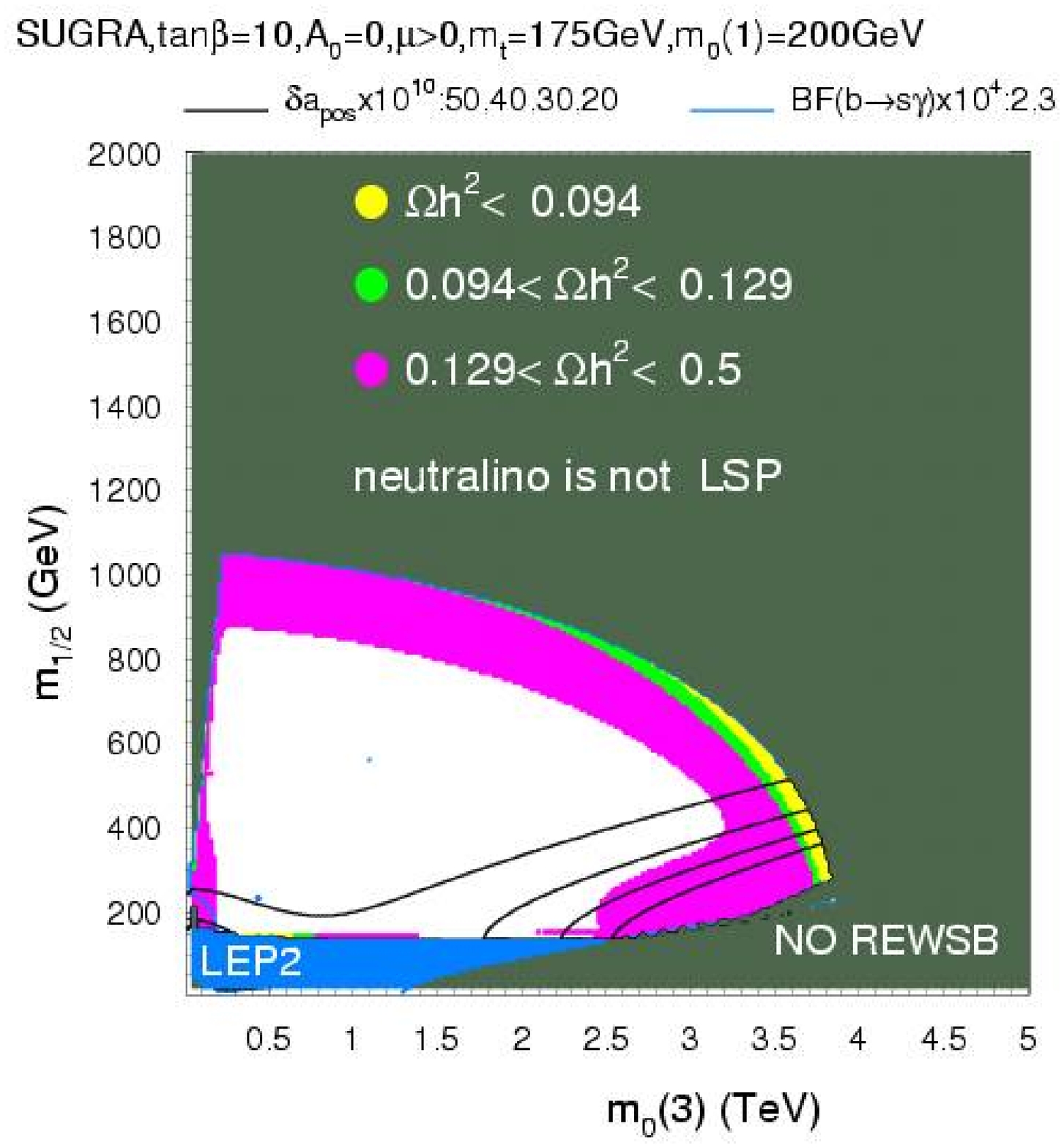,width=7cm}
\caption{Contour levels for $BF(b\to s\gamma)$, $a_\mu$ and $\Omega h^2$
in the mSUGRA model
 in the 
$m_0(3)\ vs. m_{1/2}$ plane of the NMH SUGRA model for
$m_0(1,2)=50$, 100  and 200 GeV with 
$A_0=0$, $\mu >0$, and $\tan\beta =10$.
Magenta region excluded by LEP2 searches.
}
\label{fig:nmh_10c}
}

If we shift to the $m_0(1)=100$ GeV plot, then
slepton masses correspondingly increase, suppressing the relic density 
in the low $m_{1/2}$ region of the plot. One must move close to the slepton
co-annihilation region at the upper boundary of parameter space in order 
to generate a sufficiently low value of $\Omega_{\tz_1}h^2$ in accord with
the WMAP analysis. Large $m_0(3)$ values are favored over small ones
only because small $m_0(3)$ implies larger $m_{1/2}$ values to get the
relic density right, but then the large $m_{1/2}$ value drives up the 
smuon masses and thus suppresses $\Delta a_\mu$ too much.
The further shift to $m_0(1)=200$ GeV
leads to a plot in which very little area contains a low value of 
$\chi^2$. This is due to the fact that slepton masses have increased, 
and neutralino annihilations graphs with $t$-channel slepton exchange 
are suppressed.
Only the border region survives, where $\tmu-\tz_1$ and $\te-\tz_1$ 
co-annihilation occurs at a large rate.

In Fig.~\ref{fig:nmh_30}, we again show the $m_0(3)\ vs.\ m_{1/2}$ plane,
this time for $\tan\beta =30$, and for $m_0(1)$ values of
50, 100 and 200 GeV. Fig.~\ref{fig:nmh_30c} 
presents the associated $BF(b\to s\gamma)$, $a_\mu$ and $\Omega h^2$
contour plots for the same $\tan\beta$ value.
\TABLE{
\begin{tabular}{lc}
\hline
parameter & value (GeV) \\
\hline
$M_2$ & 351.1 \\
$M_1$ & 184.2 \\
$\mu$ & 516.9 \\
$m_{\tg}$ & 1067.7 \\
$m_{\tu_L}$ & 939.8 \\
$m_{\tu_R}$ & 910.0 \\
$m_{\td_L}$ & 943.5 \\
$m_{\td_R}$ & 907.1 \\
$m_{\tst_1}$ & 1175.1 \\
$m_{\tst_2}$ & 1477.5 \\
$m_{\tb_1}$ & 1460.0 \\
$m_{\tb_2}$ & 1637.1 \\
$m_{\te_L}$ & 319.3 \\
$m_{\te_R}$ & 188.2 \\
$m_{\tnu_e}$ & 295.1 \\
$m_{\ttau_1}$ & 1386.1 \\
$m_{\ttau_2}$ & 1475.4 \\
$m_{\tnu_\tau}$ & 1468.5 \\
$m_{\tw_1}$ & 348.2 \\
$m_{\tw_2}$ & 542.4 \\
$m_{\tz_1}$ & 179.4 \\
$m_{\tz_2}$ & 347.2 \\ 
$m_A$ & 1379.3 \\
$m_h$ & 118.4 \\
$\Omega_{\tz_1}h^2$& 0.115\\
$BF(b\to s\gamma)$ & $3.52\times 10^{-4}  $\\
$\Delta a_\mu    $ & $35.1 \times  10^{-10}$\\
\hline
\end{tabular}
\caption{\label{tab:1} Masses and parameters in~GeV units
for $m_0(3),\ m_{1/2},\ A_0,\ \tan\beta ,\ sign(\mu) =$
1500~GeV, 450~GeV, 0, 30, +1 in the NMH SUGRA model.
We also take $m_H=m_0(3)$ and $m_0(1)=100$ GeV.
The spectrum is obtained using ISAJET v7.69.}
}

By increasing $\tan\beta$, we also increase 
the SUSY contribution to $a_\mu$ for a given set of slepton and
chargino/neutralino masses. The result is a band of very low $\chi^2$
points in the $m_0(1)=50$ and 100 GeV plots, where essentially
a perfect fit to $(g-2)_\mu$, $BF(b\to s\gamma )$ and $\Omega_{\tz_1}h^2$
can be obtained. The low $m_{1/2}$ portion of the $m_0(3)\ vs.\ m_{1/2}$
plane are largely excluded because they give rise to too large a relic density
of $0.129<\Omega_{\tz_1}h^2<0.5$, and because they give rise to too large
a value of $\Delta a_\mu$. In these plots, we also note that the HB/FP
strip has moved to a somewhat lower $\chi^2$ value compared to the
$\tan\beta =10$ plots due to a larger SUSY contribution to $\Delta a_\mu$.
Sample spectrum for a point with very low $\chi^2$ is shown in 
Table~\ref{tab:1}.
The last frame, with $m_0(1)=200$ GeV, has become again 
largely excluded, save for a 
narrow strip where slepton co-annihilation occurs, and in the HB/FP
region. 

%
\FIGURE{%
\epsfig{file=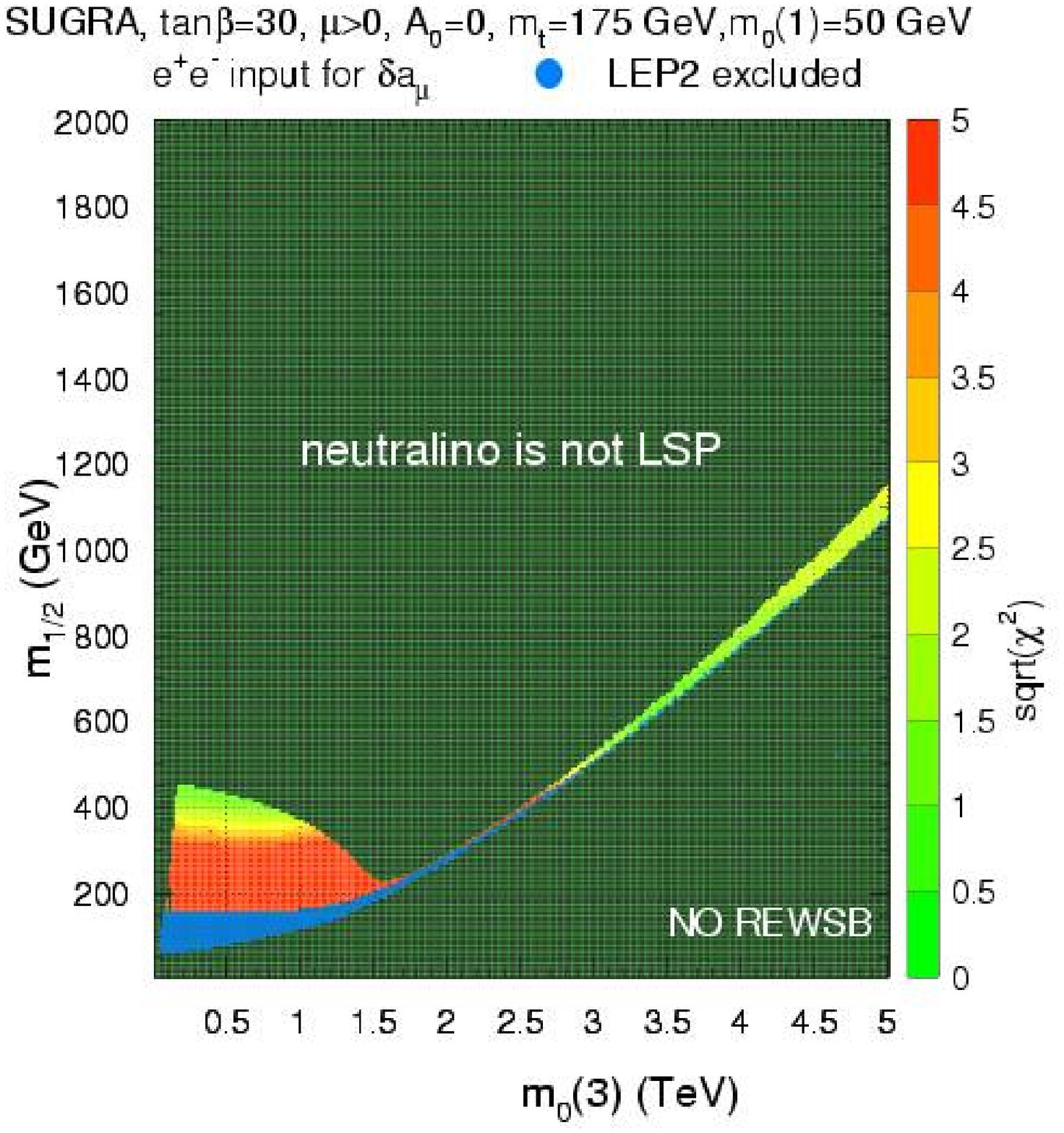,width=7cm}
\epsfig{file=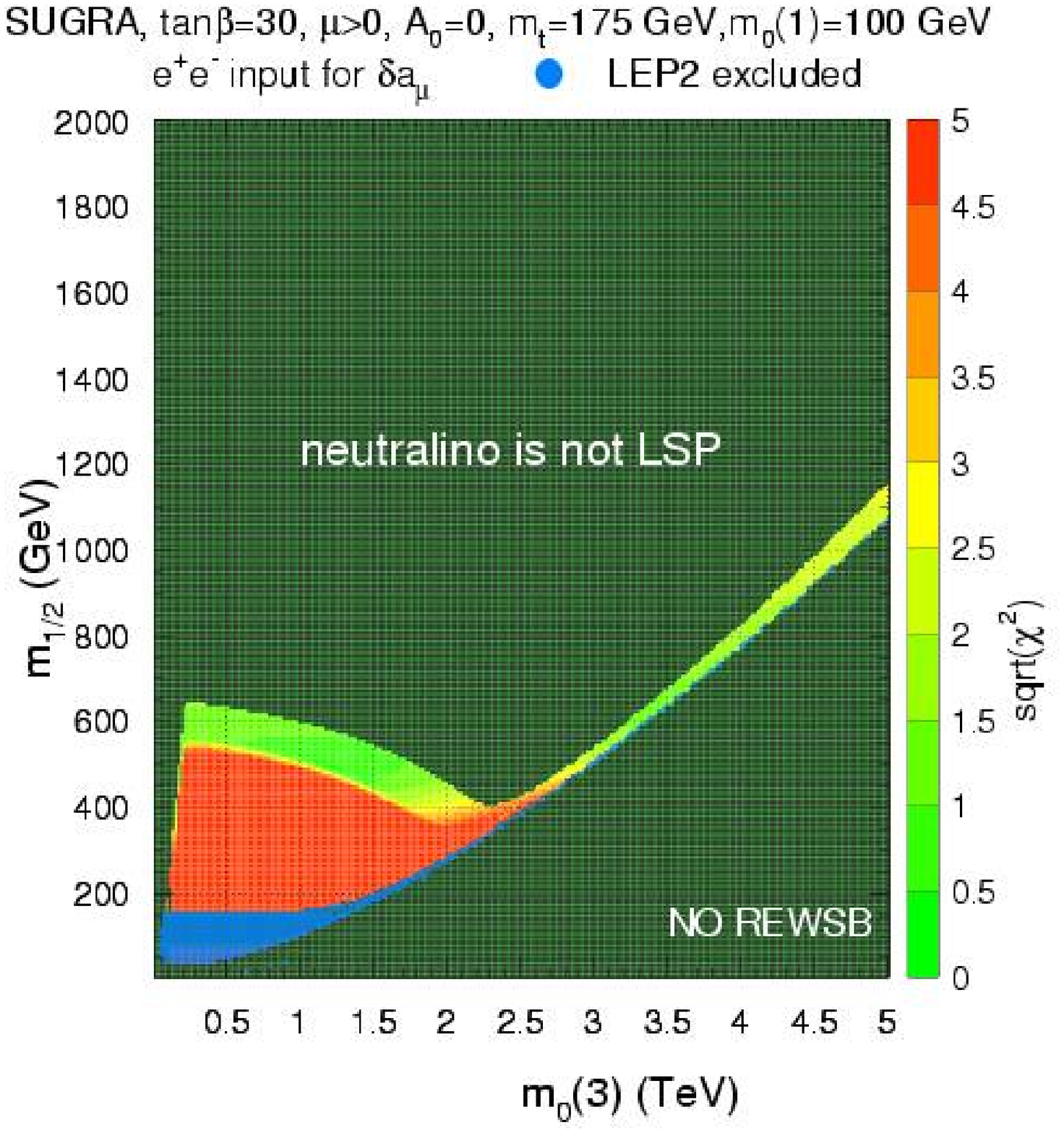,width=7cm}\\
\epsfig{file=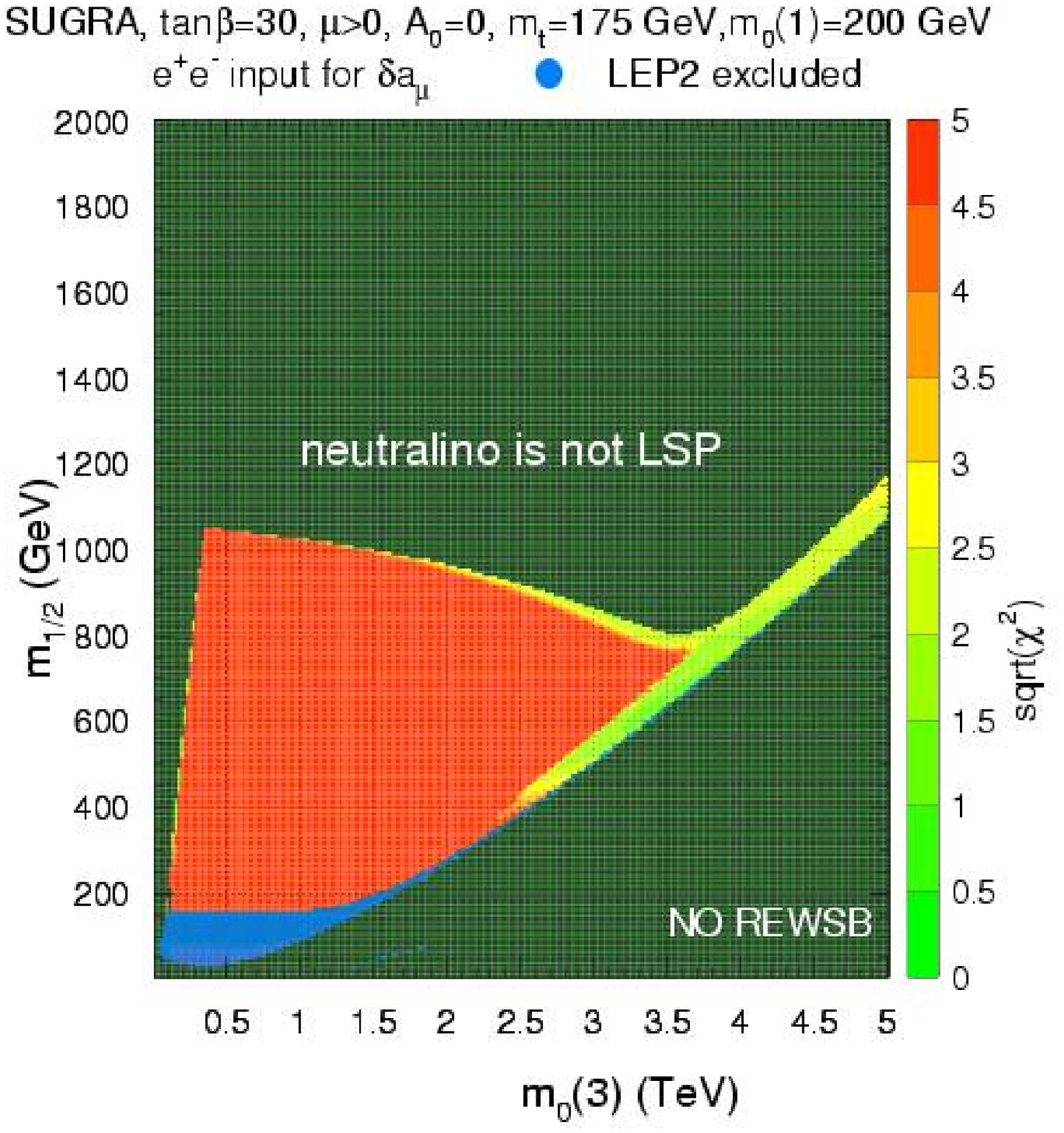,width=7cm}
\caption{Plot of regions of $\sqrt{\chi^2}$ in the 
$m_0(3)\ vs. m_{1/2}$ plane of the NMH SUGRA model for
$m_0(1,2)=50$, 100 and 200 GeV, with 
$A_0=0$, $\mu >0$, and $\tan\beta =30$.
The green regions have low $\chi^2$, while red regions have
high $\chi^2$. Yellow is intermediate.
}
\label{fig:nmh_30}
}

\FIGURE{%
\epsfig{file=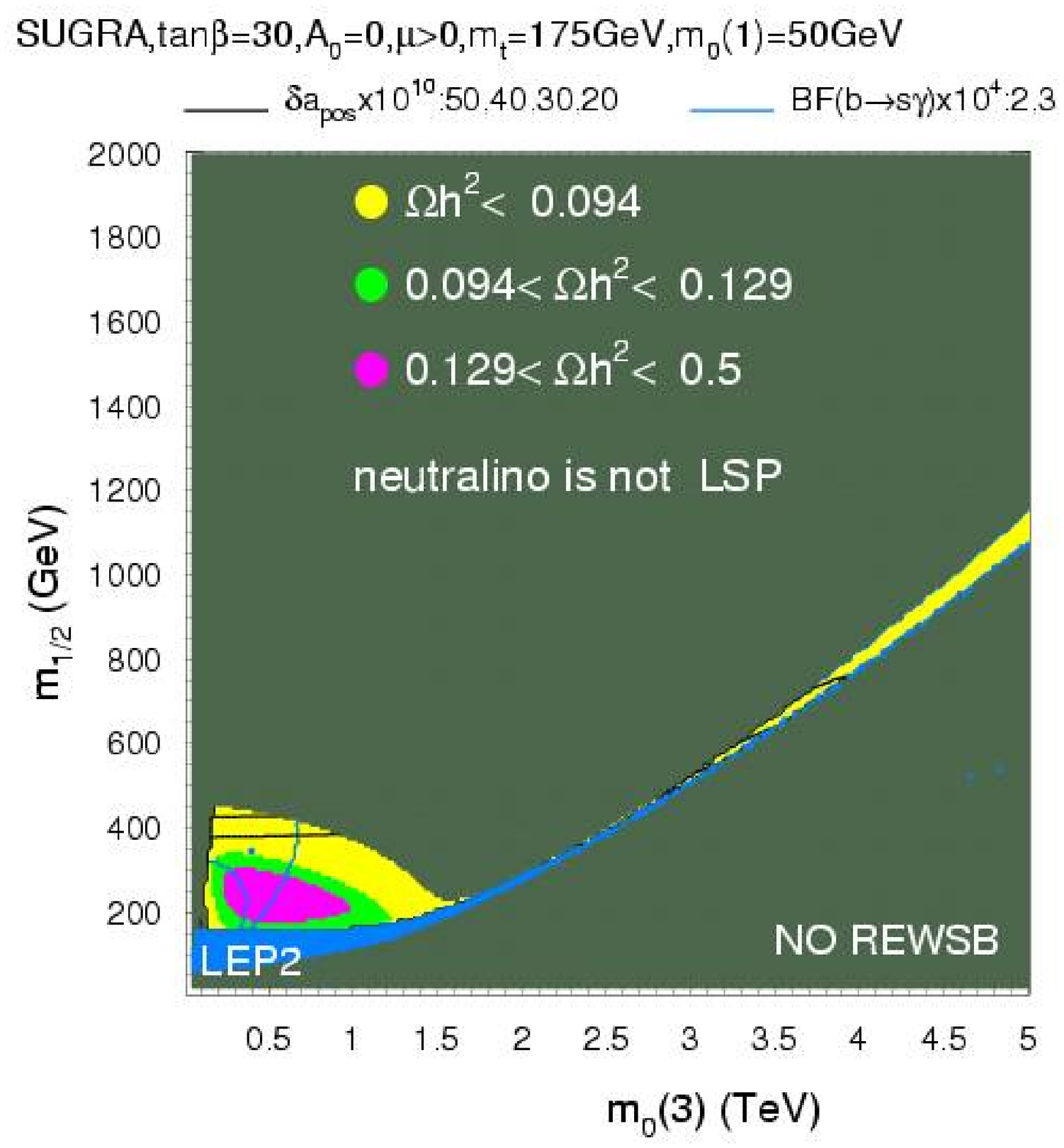,width=7cm}
\epsfig{file=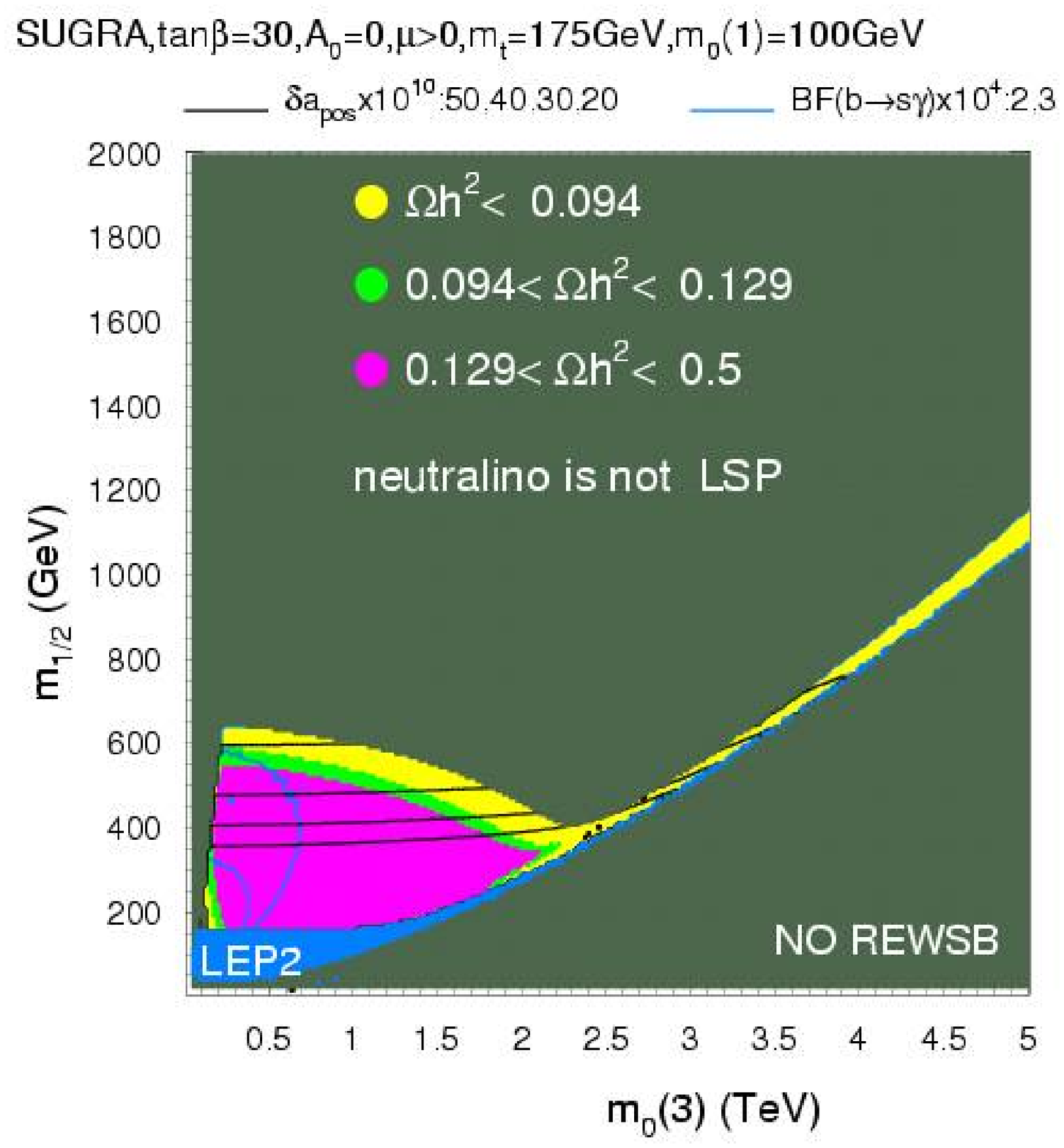,width=7cm}\\
\epsfig{file=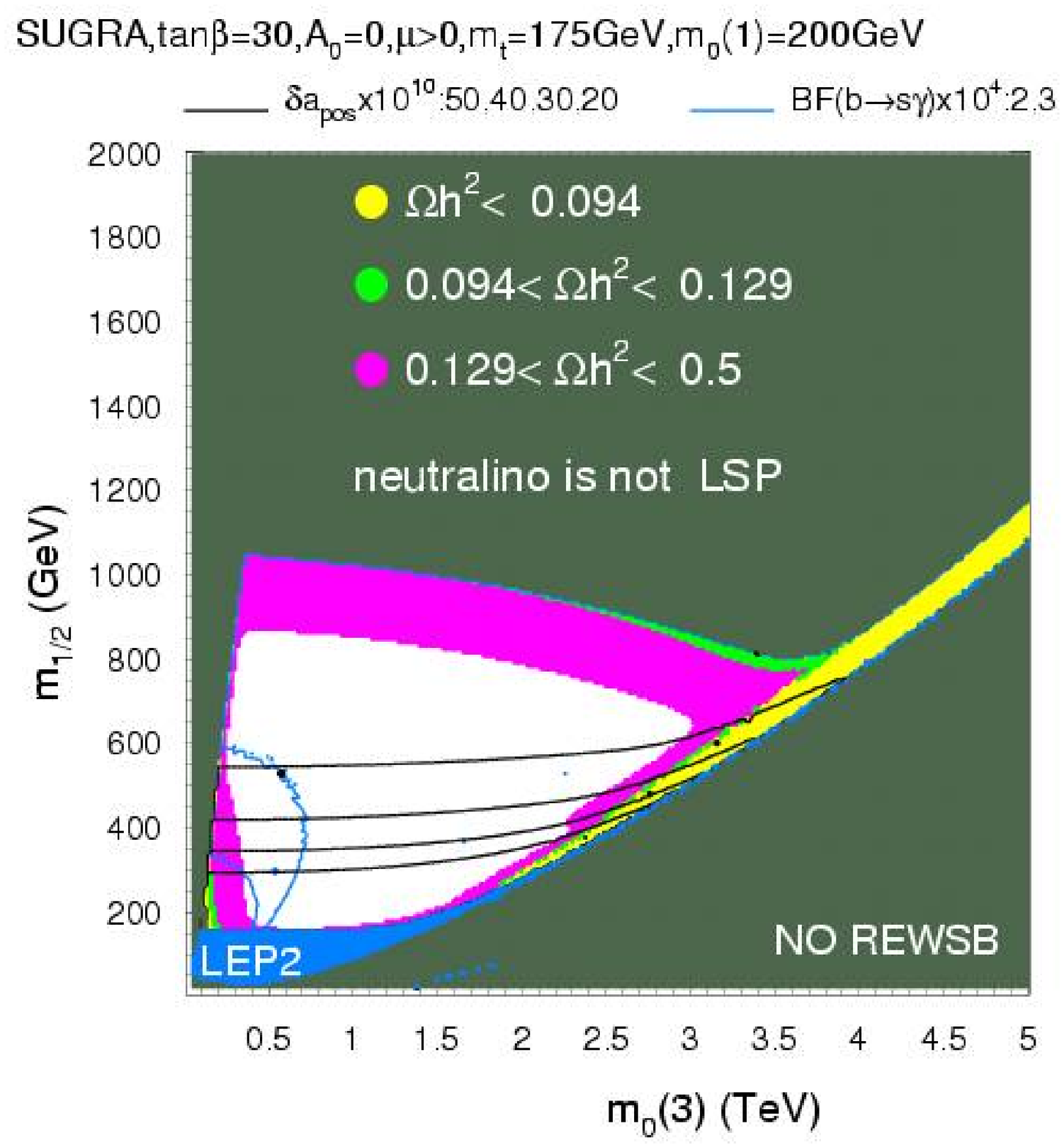,width=7cm}
\caption{
Contour levels for $BF(b\to s\gamma)$, $a_\mu$ and $\Omega h^2$
in the mSUGRA model in the 
$m_0(3)\ vs. m_{1/2}$ plane of the NMH SUGRA model for
$m_0(1,2)=50$ GeV, and $m_0(1,2)=100$ GeV with 
$A_0=0$, $\mu >0$, and $\tan\beta =30$.
Magenta region excluded by LEP2 searches.
}
\label{fig:nmh_30c}
}

%

\section{Implications for colliders}
\label{sec:collider}

In the previous section, we have seen that the adopton of non-universal
scalar masses can lead to a reconciliation of the SUSY explanation for
$a_\mu$, $BF(b\to s\gamma )$ and $\Omega_{\tz_1}h^2$. The scenario
of NMH SUGRA model works if $m_0(1)\simeq m_0(2)\ll m_0(3)$, and leads 
to spectra typically with squarks and third generation sleptons 
in the TeV range, while first and second generation sleptons 
have masses in the range of 100-300 GeV. The presence of rather
light first and second generation sleptons in the sparticle mass spectrum
in general leads to enhancements in leptonic cross sections from
superparticle production at collider experiments, compared to the case 
where selectrons and smuons are in the multi-TeV range. 
The enhancement comes from the fact that sleptons may now be produced 
with non-negligible cross sections at colliders, and also in that
their presence enhances the leptonic branching fractions of charginos
and especially neutralinos. In this section, we discuss the implications of
light selectrons and smuons for the Fermilab Tevatron collider, the
CERN LHC and a linear $e^+e^-$ collider operating with
$\sqrt{s}\sim 0.5-1$ TeV.

\subsection{Tevatron signals}

At the Fermilab Tevatron collider, the total production cross section 
for slepton pair production is rather low ($\sim 10$ fb for 
$m_{\tell}\sim 100$ GeV)\cite{bhr}, 
so that prospects for their direct detection are 
not encouraging\cite{bcpt_sl} even with optimistic projections for
the integrated luminosity to be gathered. Even so, if sleptons
are light enough, then charginos and neutralinos may directly decay 
into them via two body modes: $\tw_1\to \nu_\ell\tell_L,\ \tnu_\ell \ell$
and $\tz_2\to \tell_R\ell$, $\tnu_\ell\nu_\ell$ and $\tell_L\ell$ 
(we suppress bars on anti-particles).
Even if two-body decays are not kinematically open, then relatively light 
sleptons can yield enhancements in chargino and especially neutralino
three body decay to leptons.

If sparticles are accessible to the Fermilab Tevatron, then it is 
usually expected that $p\bar{p}\to \tw_1^+\tw_1^- X$ and
$\tw_1\tz_2 X$ will be the dominant production cross sections\cite{tev_susy}.
If $\tw_1\to\ell\nu_\ell\tz_1$ and $\tz_2\to \ell\bar{\ell}\tz_1$, then
clean trilepton signals may occur at an observable rate\cite{3l}. Signal
and background rates have recently been investigated in Ref. \cite{trilep}, 
where Tevatron reach plots may also be found. Here, we generate all
sparticle production processes using Isajet 7.69 for the
parameter space point $m_0(3)=1400$ GeV, $m_{1/2}=225$ GeV, $A_0=0$, 
$\tan\beta =10$ and $\mu >0$. We plot the isolated trilepton signal
after using cuts SC2 of Ref. \cite{bdpqt}, where the backgrounds are also
evaluated. The results, plotted versus variation in the $m_0(1)$ parameter,
are shown in Fig. \ref{fig:tevatron}, where the signal level needed
for a $5\sigma$ signal with 10 fb$^{-1}$ is also denoted. The error bars show 
the Monte Carlo statistical error. When 
$m_0(1)=m_0(3)$ (at $m_0(1)=1400$ GeV), the results correspond to the mSUGRA
model, and the isolated trilepton signal is well below discovery
threshold. As $m_0(1)$ decreases to smaller values, the isolated 
trilepton rate drops. This is due in part to a slight 
reduction in total SUSY cross section (shown in frame {\it b})), but also 
due to an interference in the $\tz_2$ leptonic decay rates, 
due to destructive interference between slepton and $Z$ boson mediated graphs.
As $m_0(1)$ drops to even lower values, the light sleptons begin to dominate
neutralino three body decays rates, and consequently 
the trilepton cross section rises steeply, to the level of observability. 
Eventually chargino and and neutralino two body decays 
to sleptons turn on (in this case, first $\tz_2\to \tell_R\ell$), 
and trilepton rates become very high. For even
lower $m_0(1)$ values, neutralino decays to $\tnu_\ell\nu_\ell$ turn on,
and briefly suppress the trilepton rate, until finally
$\tz_2\to\tell_L\ell$ turns on, and the trilepton rate picks up again for the
lowest $m_0(1)$ values.
\FIGURE{\epsfig{file=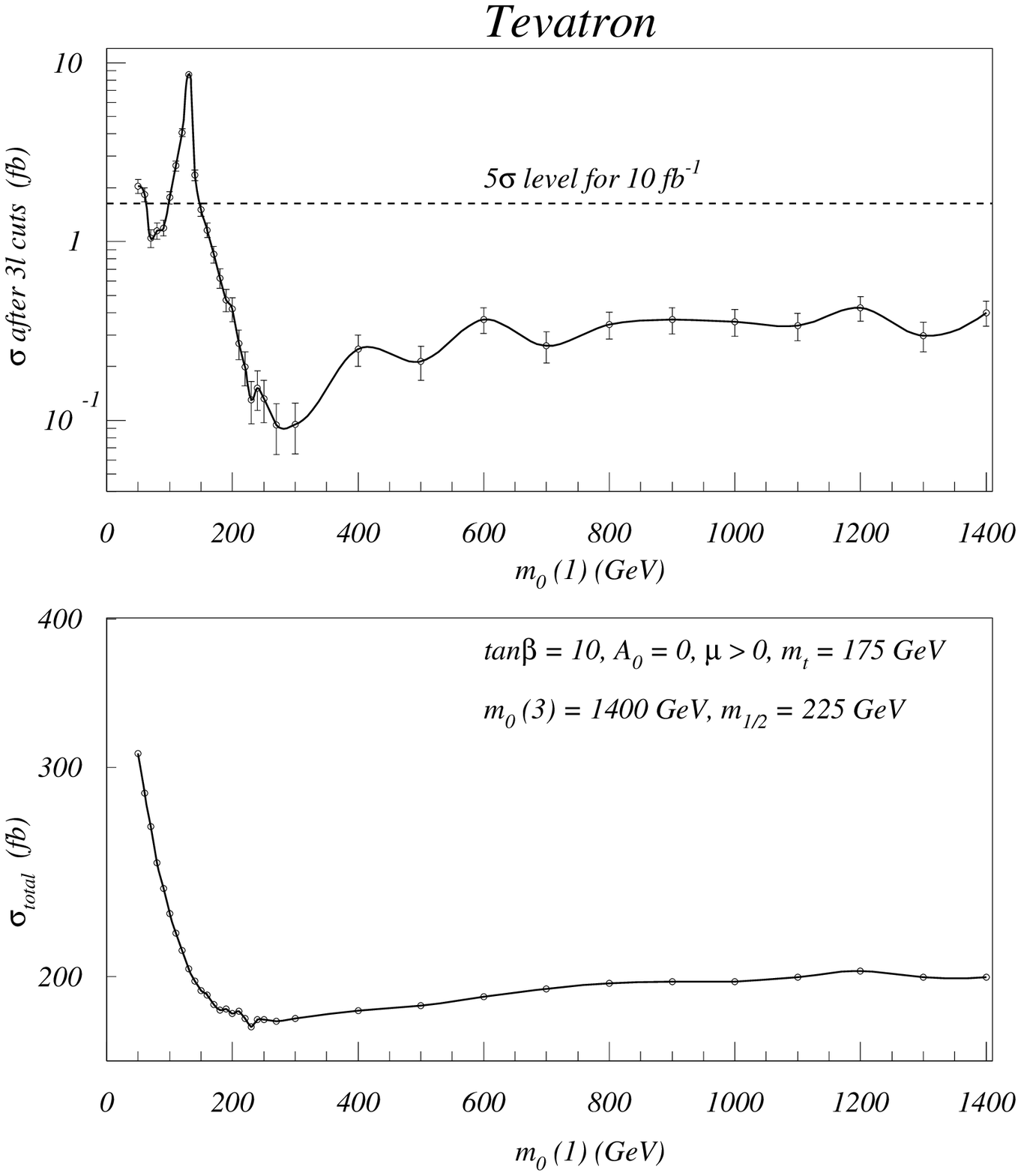,width=10cm}
\caption{Rates for isaolated trilepton events at the Fermilab Tevatron
$p\bar{p}$ collider, after cuts SC2 from Ref. \cite{trilep}.
}
\label{fig:tevatron}
}

The large rates for clean trileptons at the Fermilab Tevatron occur for 
$m_0(1)$ values below about 200 GeV, which is just the regime needed
to give substantial SUSY contributions to $(g-2)_\mu$. In the plot shown, 
the $m_{1/2}$ value chosen was small enough that sparticle total production 
cross sections are large enough to generate an observable signal. For
larger $m_{1/2}$ values, an observable signal at the Fermilab Tevatron 
doesn't always occur, even if $m_0(1)$ is small.

\subsection{CERN LHC}

The reach of the CERN LHC for SUSY particles in the mSUGRA model
extends up to $m_{1/2}$ values as high as 1400 GeV (700 GeV) for
small (large) values of the parameter $m_0$, assuming 100 fb$^{-1}$
of integrated lu\-mi\-no\-si\-ty\cite{lhcreach}. This corresponds to a reach in 
$m_{\tg}$ of 3 (1.8) TeV, respectively.
For the case of the NMH SUGRA model, $m_{1/2}$ is bounded from above by the
requirement that the neutralino, and not a slepton, is the LSP. 
The upper bound on $m_{1/2}$ is well below the LHC reach in $m_{1/2}$ for 
the mSUGRA model, and so the CERN LHC should easily establish a 
signal in all of the NMH SUGRA model parameter space with 
low enough $m_0(1)$ values as to
satisfy the $(g-2)_\mu$ deviation. A possible exception occurs if SUSY
lies in the HB/FP region. But even here, first and second generation 
sleptons are relatively light, and may be accessible to LHC searches in 
the dilepton channel\cite{lhc_sleptons}.

The collider signals for SUGRA-like models at the CERN LHC are naturally 
divided up according to number of leptons in the final state. Thus, 
in Ref. \cite{lhcreach}, signals for jets plus $\eslt$ plus 0, 1, 
2 same sign (SS) or 2 opposite sign leptons (OS), 3 leptons and 4 or more
leptons occur. In Fig. \ref{fig:lhc}, we plot signal rates from the
NMH SUGRA model for $m_0(3)=2300$ GeV, $m_{1/2}=300$ GeV, 
$A_0=0$, $\tan\beta =10$ and $\mu >0$, versus $m_0(1)$. 
The cuts are listed in Ref. \cite{lhcreach}, and have been optimized to give
best signal-to-background ratio for $m_0(1)=100$ GeV. For very large
$m_0(1)\simeq m_0(3)$ (mSUGRA model case), a variety of nonleptonic and 
multilepton signals occur, all at observable levels above background
estimates. As $m_0(1)$ decreases, the total 
sparticle production cross section increases substantially, mainly because
first and second generation squarks are decreasing in mass, and enhancing
strongly interacting sparticle pair production rates.
As $m_0(1)$ decreases further, leptonic rates also increase, in part because of increased total production cross sections, but also due to 
enhanced chargino and neutralino (s)leptonic branching fractions.
As $m_0(1)$ drops below about 200 GeV (the value typically needed to 
explain the $(g-2)_\mu$ anomaly), multilepton rates rise steeply.
Thus, we would expect that SUSY as manifested in the NMH SUGRA model
would be easily discovered, and what's more, the signal events would be
unusually rich in multilepton events. Such multilepton events can be 
especially useful for reconstructing sparticle masses in gluino and squark
cascade decay events\cite{frank}.
\FIGURE{\epsfig{file=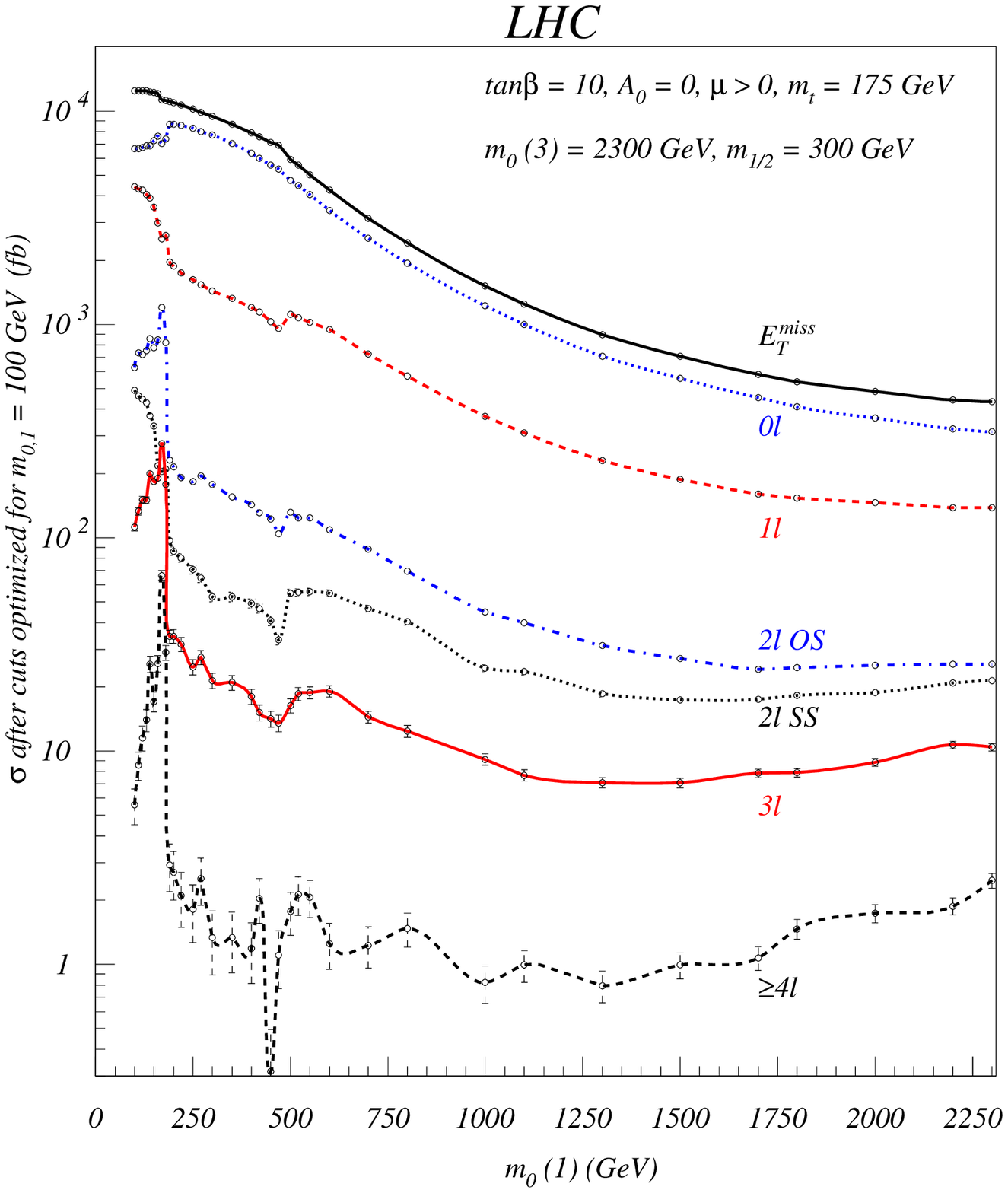,width=10cm}
\caption{Rates for various SUSY signals 
at the CERN LHC, using cuts from Ref. \cite{lhcreach}.
The $5\sigma$ discovery cross section for the $\eslt +$ jets channel for
100 fb$^{-1}$ of integrated luminosity is $53.4$ fb.
}
\label{fig:lhc}
}

\subsection{$e^+e^-$ linear collider}

A linear collider operating at $\sqrt{s}=0.5-1$ TeV may be the next frontier
particle physics accelerator beyond the CERN LHC. 
Depending on sparticle masses and the collider energy, 
charginos and neutralinos may or may not be accessible. However, 
in the NMH SUGRA model, light first and second generation sleptons are 
needed both to explain the $(g-2)_\mu$ anomaly, but also to enhance
neutralino annihilation in the early universe. This means slepton masses
are typically in the 100-300 GeV range, and likely within reach of 
a linear $e^+e^-$ collider\cite{nlcreach}.

\section{Conclusions}
\label{sec:conclusions}

In this paper, we have examined the parameter space of 
gravity mediated SUSY breaking models, to see if constraints from
$(g-2)_\mu$, WMAP $\Omega_{\tz_1}h^2$ and $BF(b\to s\gamma )$ can
simultaneously be satisfied. In performing this task, we used new
theoretical and experimental determinations of $(g-2)_\mu$ which make use of
$e^+e^-\to hadrons$ data at low energy to fix the hadronic vacuum
polarization contribution to $a_\mu =(g-2)_\mu$, 
and which gives a $\sim 3\sigma$
discrepancy with the SM.
While $(g-2)_\mu$ favors a spectrum of relatively light second generation
sleptons, the $BF(b\to s\gamma )$ measurement favors relatively
heavy (TeV scale) third generation squarks. 
Satisfying all constraints is possible in the mSUGRA model in only very 
limited regions: the stau co-annihilation region at low $\tan\beta\sim 10$, 
and in the $A$-annihilation funnel at very large $\tan\beta$. In particular, 
the HB/FP region gives only a tiny contribution to $(g-2)_\mu$, and is
disfavored by about $3\sigma$ if the $e^+e^-$ data is used for the 
hadronic vacuum polarization determination.

We advocate relaxing the universality assumption between 
third generation scalars and their first and second generation
counterparts.
Simultaneously fulfilling all three constraints favors low values
of $m_0(1)\simeq m_0(2)\sim 50- 200$ GeV, while $m_0(3)$ values of a TeV or 
beyond are preferred. 
This comprises a normal scalar mass hierarchy, or NMH, at the GUT scale.
The weak scale mass splittings amongst the squarks
can be modest in NMH SUGRA scenario, even though
mass splittings at $M_{GUT}$ are large.
The FCNC constraints from $B_d-\overline{B}_d$  and $BF(b\to s \gamma)$
are very weak and hardly affect the NMH SUGRA model parameter space.
However, the weak scale splitting amongst sleptons is large, 
with selectrons and smuons in the $100-300$ GeV range, while staus are 
in the TeV range. Neutralinos in the early universe annihilate via a 
combination of $t$-channel slepton exchange and slepton co-annihilation.

A testable consequence of the NMH scenario is that sleptons may be 
directly observable at the CERN LHC, provided that the slepton-neutralino
mass gap isn't too small. In addition, the presence of 
relatively light sleptons enhances the rates for multilepton
production in SUSY cascade decay events at the CERN LHC. The light sleptons 
might also lead to observable rates for trilepton events at the Fermilab 
Tevatron collider. Finally, selectrons, smuons and their associated
sneutrinos in the $100-300$ GeV mass range should be
acccessible to a linear $e^+e^-$ collider operating at $\sqrt{s}=0.5-1$ TeV, 
even if squarks and third generation sleptons are beyond the limit for
direct searches.

\acknowledgments
 
We thank X. Tata for discussions. 
A.B. would like to thank David Hertzog for very usefull
and stimulating discussions at ASPEN03 conference.
This research was supported in part by the U.S. Department of Energy
under contract number DE-FG02-97ER41022.

\section*{Appendix}

In this appendix, we decribe the 
``minimal mixing'' scenario for soft SUSY breaking mass terms
we adopted for our study, and apply it to constraints from
$BF(b\to s\gamma )$\cite{fcnc}.
(Our notation is similar to that of \cite{Misiak:1997ei}.)
The relevant Lagrangian terms are
\be
{\cal L}\ni -\tilde Q_i^\dagger ({\bf m}^2_Q)_{ij} \tilde{Q}_j
+\left( 
({\bf a}_d)_{ij}\tilde{Q}_i^a H_{da}\tilde{d}_{Rj}^\dagger +h.c.\right),
\ee
where $i,j=1-3$ are generation indices and $a$ is an $SU(2)$ index.
We will ignore the possibility of large $CP$ violating phases 
for this sample analysis.

Constraints are generally presented in terms of soft SUSY breaking matrix 
elements in the super-CKM basis, where quark mass matrices are
diagonalized by
\be
{\bf m}_u=v_u\ V_R^u{\bf f}_u^T V_L^{u\dagger}\ \ {\rm and}\ \ 
{\bf m}_d= v_d\ V_R^d{\bf f}_d^T V_L^{d\dagger},
\ee
and where the CKM matrix is given by $K\equiv V_L^u V_L^{d\dagger}$.
We adopt a basis wherein ${\bf f}_u$ is already diagonal, so that
$V_L^{u\dagger}=V_R^u={\bf 1}$ and $K=V_L^{d\dagger}$.

At $Q=M_{GUT}$ the SSB mass-squared matrix has the form
\be
{\bf m}_Q^2=\left[\begin{array}{ccc} m_0^2(1) & 0 & 0\\
0 & m_0^2(1) & 0\\ 0 & 0 & m_0^2(3) \end{array}\right] ,
\ee
while the trilinear matrix ${\bf a}_d$ has only one non-zero entry
given by $({\bf a}_d)_{33}=f_b A_0$, where $f_b$ is the $b$-quark 
Yukawa coupling.

In the minimal mixing scenario,
the generation of off-diagonal terms in the
running from $M_{GUT}$ to $M_{weak}$ is neglected.
Therefore all off-diagonal
elements in the down type squark mass matrix sub-blocks
$({\bf m}_{\td}^2)_{LL}$ and $({\bf m}_{\td}^2)_{LR}$ 
arise from the 
rotation to the super-CKM basis. 
These off-diagonal elements are of the order
$(\Delta_d)_{LL} \alt 10^{-2}\times (m^2_{d1})_{LL}$
while
$(\Delta_d)_{LR} \alt 10^{-4}\times (m^2_{d3})_{LL}$.
For example,
for our sample  parameter space point
given by (\ref{eq:sample})
these matrices will have the following numerical form 
\be
({\bf m}_{\td}^2)_{LL}
\simeq\left[\begin{array}{ccc} 1.56\times 10^6 & -39.3 & 2550\\
 -39.3 & 1.56\times 10^6 & -1.8\times 10^4 \\ 2550 & -1.8\times 10^4 
& 2.0\times 10^6 \end{array}\right] ,
\ee
and
\be
({\bf m}_{\td}^2)_{LR}\simeq\left[\begin{array}{ccc} -0.1 & 0.9 & -22.2\\
 0.9 & -6.6 & 160.3 \\ -22.2 & 160.3 & 
-3909 \end{array}\right] ,
\ee
where all entries are in units of GeV$^2$.

It is also possible to go a step further than the minimal mixing scenario
and include (two-loop) RG evolution of SSB matrices 
${\bf m}_Q^2$ and ${\bf a}_d$ from $Q=M_{GUT}$ to $Q=M_Z$ to  
generate flavor violating soft terms.
We have checked using SoftSUSY\cite{softsusy} that these off-diagonal elements
would be of the same order of magnitude
as off-diagonal elements of SSB matrices in the minimal mixing scenario
after rotation to super-CKM basis.
Therefore, for our order-of-magnitude estimation
the minimal mixing scenario appears justified.
For example,
for our sample  parameter space point (\ref{eq:sample}),
the SSB matrices after RG running from $M_{GUT}$ to $M_Z$
will have the following numerical form:
\be
{\bf m}_Q^2\simeq\left[\begin{array}{ccc} 1.56\times 10^6 & -43 & -660\\
 -43 & 1.56\times 10^6 & -7625 \\ -660 & -7626 & 
2.0\times 10^6 \end{array}\right]
\ee
(units of GeV$^2$), and
\be
{\bf a}_d\simeq \left[\begin{array}{ccc} -2.1 & -4.1 & -2.3\\
 -4.1 & -20.0 & -28.1 \\ -2.2 & -25.9 & 
-665.1 \end{array}\right] 
\ee
(in units of GeV).

A rotation of the down squark sub-block mass matrices 
to the super-CKM basis yields
\be
({\bf m}_{\td}^2)_{LL}\simeq\left[\begin{array}{ccc} 1.56\times 10^6 & -164 & 3601\\
 -164 & 1.56\times 10^6 & -2.6\times 10^4 \\ 3601 & -2.6\times 10^4 
& 2.0\times 10^6 \end{array}\right] ,
\ee
and
\be
({\bf m}_{\td}^2)_{LR}\simeq\left[\begin{array}{ccc} -6.7 & 0.04 & 1.1\\
 0.04 & -116.4 & -8.4 \\ -0.66 & 4.2 & 
-3922.8 \end{array}\right] .
\ee
In this example at least, it is clear
that the off-diagonal elements after including RG 
evolution are of the same order
of magnitude as those in the minimal mixing scenario.

One should also notice that $(\Delta_d)_{LL}$
is proportional to $m^2_{d3}-m^2_{d1}$,
which in our scenario is of the order $({\rm TeV})^2$, 
while the $(\Delta_d)_{LR}$
elements are at most of the order $(10\ {\rm GeV})^2$ and therefore 
contributions from $A$ terms can be neglected in our 
estimates of constraints from $\Delta m_K$ and $\Delta m_B$.
 

The general bounds from $BF(b\to s\gamma )$ on these matrix elements 
from \cite{fcnc} are :
\be
(\Delta_d^{23})_{LL}<8.2\left(\frac{m_{\tq}}{500\ {\rm GeV}}\right)^2 m_{\tq}^2
\ee
which 
for $(\Delta_d^{23})_{LL}\sim 10^{-2}\times m_{\tq}^2$  
implies  $m_{\tq}> 17.5$~GeV, which
is easily satisfied.
Also,
\be
(\Delta_d^{23})_{LR}<0.016\left(\frac{m_{\tq}}{500\ {\rm GeV}}\right)^2 m_{\tq}^2
\sim 64\ {\rm GeV}^2
\ee
for $m_{\tq}\sim 1$ TeV.
Thus, the assumed mass splittings in the NMH scenario seem
safe from FCNC constraints when minimal mixing is assumed.

%

\end{document}